\documentclass[%
 reprint,
amsmath,amssymb,
aps,
]{revtex4-2}

\usepackage{graphicx}
\usepackage{dcolumn}
\usepackage{bm}
\usepackage{xcolor}
\definecolor{darkblue}{rgb}{0,0,0.6}
\usepackage[colorlinks,linkcolor=darkblue,citecolor=darkblue,urlcolor=darkblue]{hyperref}

\usepackage{physics}
\usepackage[separate-uncertainty = true, number-unit-product = \;]{siunitx}
\usepackage{bbm}

\newcommand{\mb}[1]{\mathbf{#1}}
\newcommand{\mrm}[1]{\mathrm{#1}}

\def\evaluateaux#1#2{{#1\,\smash{\vrule height .8\ht1 depth .85\dp1}}_{\,#2}}
\def\evaluate#1#2{\mathchoice
  {\setbox1\hbox{${\displaystyle #1}_{\scriptstyle #2}$}
  \evaluateaux{#1}{#2}}
  {\setbox1\hbox{${\textstyle #1}_{\scriptstyle #2}$}
  \evaluateaux{#1}{#2}}
  {\setbox1\hbox{${\scriptstyle #1}_{\scriptscriptstyle #2}$}
  \evaluateaux{#1}{#2}}
  {\setbox1\hbox{${\scriptscriptstyle #1}_{\scriptscriptstyle #2}$}
  \evaluateaux{#1}{#2}}}

\begin{document}

\title{Nucleation beyond Equilibrium: Fronts Control Invasion in Bistable Ecosystems}

\author{Victor Lequin}
\email{victor.lequin@phys.ens.fr}
\affiliation{
 Laboratoire de Physique de l'Ecole normale supérieure, ENS, Université PSL, CNRS, Sorbonne Université, Université de Paris, F-75005 Paris, France
}
\author{Giulio Biroli, Camille Scalliet}
\affiliation{
 Laboratoire de Physique de l'Ecole normale supérieure, ENS, Université PSL, CNRS, Sorbonne Université, Université de Paris, F-75005 Paris, France
}

\date{\today}

\begin{abstract}
Bistability, the existence of two alternative stable states with distinct basins of attraction, is common across ecology and many other biological, chemical, and physical systems. In spatially extended systems, the invasion of one state by the other proceeds through nucleation, the fluctuation-driven growth of a droplet beyond a critical size. Classical Nucleation Theory (CNT) quantifies this process using the energy landscape, but this description relies on detailed balance, the condition of microscopic reversibility that holds at equilibrium. Ecological dynamics generally violate detailed balance and are described not by a single scalar field but by several coupled, non-conserved abundances--a vector order parameter--for which no general nucleation theory exists. Here we derive such a theory for reaction-diffusion systems with a non-conserved vector order parameter, valid both close to the binodal, where invasion proceeds through propagating fronts, and close to the spinodal, where the metastable state loses stability. Extensive numerical computations of the quasipotential confirm the theory in both regimes. We show that the mathematical structure of CNT survives out of equilibrium, once energetic quantities are replaced by dynamical properties of fronts: the front speed plays the role of the bulk free-energy difference between phases, and the diffusivity that of the surface tension. Applied to the two-species Lotka-Volterra model, an archetypal system of bistable ecological antagonism, our theory shows that strong interspecific competition generates a pronounced depletion region within fronts, where the total abundance falls well below carrying capacity. This vectorial structure, invisible to a scalar description based on species frequency alone, makes invasion exponentially harder as competition strengthens at fixed competitive advantage--a prediction testable in microbial systems.
\end{abstract}

\maketitle

\section{Introduction}

What fate awaits an alien species introduced into an environment dominated by a resident one? Whether it dies out or takes over depends on the size of its inoculum, when the two species compete strongly or express strong antagonism. Success means the invader displaces the resident, settling the community into a new stable state. A system able to sustain two such alternative stable states is called bistable. Multistability is common in ecology~\cite{ros_generalized_2023, lopes_cooperative_2024}, and bistability also arises in many biological~\cite{ferrell_self-perpetuating_2002, tan_emergent_2009, pettit_estell_population-level_2025}, chemical~\cite{russew_photoswitches_2010, wilhelm_smallest_2009} and physical~\cite{coleman_aspects_1985, debenedetti_metastable_2020, clark_di_leoni_phase_2020} systems at disparate scales. Its generality stems from the simplicity and the universality of the phase portrait that characterizes it, consisting of two basins of attraction separated by a frontier, the separatrix~\cite{strogatz_nonlinear_2019, kuznecov_elements_2023}.

In a well-mixed community, bistability is well understood: either competitor invades whenever its inoculum exceeds a critical frequency~\cite{mcnally_killing_2017}. Space changes this picture. In metacommunities, which consist of many local communities coupled by dispersal, experiments on genetically engineered yeast strains have shown that an inoculum must instead overcome a critical \emph{radius} to grow~\cite{giometto2021}. The same behavior is predicted when bistability arises from a strong Allee effect in a single species~\cite{lewis_allee_1993}. The spatial growth of the inoculum is classically described by invasion fronts, which correspond to profiles of abundances across the transition between regions dominated by each species. Fronts have been studied in depth when their profile is captured by a single scalar variable~\cite{murray_mathematical_1989, cross_pattern_1993, vansaarloos_front_2003}, such as the local frequency of one species. 

This deterministic account is only one facet of a more complex phenomenon. In fact, ecological dynamics are typically noisy--noise may be demographic, from the randomness of individual births and deaths, or environmental, from fluctuations of fitness~\cite{engen_demographic_1998}. Noise has a major effect on threshold phenomena such as the existence of a critical frequency or a critical radius: it may allow, by chance, an inoculum smaller than the threshold to grow and establish.

In physics, exactly this combination--a critical droplet in a spatially extended bistable system, together with its growth beyond a critical size by noise--constitutes the phenomenon of nucleation. There, the critical size and the nucleation probability are quantified by energetic arguments~\cite{debenedetti_metastable_2020, cahn_free_1959, langer_theory_1967, langer_statistical_1969}. These arguments rest, however, on detailed balance, the condition of microscopic time-reversibility characteristic of equilibrium systems, where nucleation is most commonly studied. Ecological systems, on the contrary, generically break detailed balance: the interactions that govern their behavior need not derive from a potential. Noise is typically state-dependent, so no energy function is available to make predictions.

How nucleation can be described without detailed balance is largely an open question. The existing out-of-equilibrium theories developed for active matter systems \cite{cates_classical_2023, zakine_unveiling_2024}\ all describe a scalar order parameter: a single field that distinguishes the two coexisting states.  The ecological setting faces two main challenges. First, as for all other problems of nucleation without detailed balance, with no energy landscape to read it from, metastability can no longer be inferred from a potential ; it must be determined dynamically. As we shall show, this can be done using the properties of invasion fronts. Second, the customary reduction to a scalar order parameter presupposes that the total abundance is uniform across space, and only the relative frequency of the two species changes spatially~\cite{lavrentovich_asymmetric_2014, tanaka_spatial_2017, lavrentovich_nucleation_2019}. We find that this presupposition breaks down: along a front, the total abundance can fall well below the carrying capacity, forming a \emph{depletion region}, already visible in the yeast experiments~\cite{giometto2021}. Capturing it requires promoting the abundances to a \emph{vector} order parameter, a set of local fields characterizing each species' abundance, whose structure inside fronts governs their motion.

In this work, we derive a theory of nucleation for systems described by a vector order parameter that may be driven out of equilibrium, either because the local flow is not the gradient of a potential or because of multiplicative (state-dependent) noise. Our framework provides a general theoretical setting for nucleation in bistable reaction-diffusion systems. We apply our theory to the prototypical two-species Lotka--Volterra competition model, which serves both as a relevant test for our theory and as a worthy problem in itself, providing valuable ecological insight.

Our general conclusion is that, close to the \textit{binodal}, where both states are equally stable, so that fronts are slow and critical droplets large, the formulas governing nucleation at equilibrium known as Classical Nucleation Theory (CNT)~\cite{langer_theory_1967,Oxtoby_1992,debenedettibook,karthika_review_2016} remain valid out of equilibrium, provided the energetic quantities appearing in them are replaced by dynamical quantities characterizing fronts. There are cases in which fronts do not suffice: close to the \emph{spinodal} transition, defined as the loss of stability of the metastable state, fronts can no longer provide the answer. We instead calculate the nucleation probability by extending the physical theory of spinodal nucleation out of equilibrium.

Applied to the Lotka--Volterra model of ecological antagonism, the theory quantifies the probability that a circular inoculum of arbitrary radius invades. This probability is related to the internal structure of vector fronts, which we show matters ecologically. When antagonism is strong, the depletion region is pronounced: the total abundance drops sharply inside the front, which markedly slows it, enlarges the critical radius, and lowers the invasion probability for a given competitive advantage. Our theory, which allows to quantify this effect, predicts that strongly competing communities are thus {\it exponentially} harder to invade than weakly competing ones at equal competitive advantage--an effect invisible to a single-frequency description~\cite{lavrentovich_asymmetric_2014, tanaka_spatial_2017, lavrentovich_nucleation_2019}, and which is testable in microbial systems.

A characteristic feature of ecological systems is that their order parameter is not conserved, a direct consequence of births and deaths, in contrast to the density in phase-separating fluids. A theory of nucleation for a conserved, scalar order parameter, such as the density in active liquids, was developed recently~\cite{cates_classical_2023}. While this article was being written, we were made aware of a concurrent work describing nucleation for non-conserved, scalar order parameters~\cite{chatzittofi_nonequilibrium_2026, ziethen_nucleation_2026}; our framework instead treats the vector case, and derives the definition of the interface position as a solvability condition rather than positing it. We discuss similarities and differences in Section~\ref{sec:conclusion}.

In Section~\ref{sec:mainresults}, we introduce the scope of our theory and present our main results, first in the general reaction-diffusion setting, and then in the Lotka--Volterra (LV) ecological one. 
In Section~\ref{sec:nuceq} we recall known results on nucleation for equilibrium dynamics and introduce the large-deviation tools used to compute the nucleation probability. In Section~\ref{sec:noneqnuc} we derive our general theory of nucleation out of equilibrium. In Section~\ref{sec:num} we present the numerical methods used to test the theory. The numerical and theoretical results as well as their ecological implications for the LV model are presented in Section~\ref{sec:frontslv} for its fronts and Section~\ref{sec:applicationlv} for its nucleation properties. We discuss broader implications and perspectives in Section~\ref{sec:conclusion}.

Section~\ref{sec:mainresults} provides a self-contained overview of the scope and main results of the paper.

\section{Main Questions and Results\label{sec:mainresults}}

\begin{figure*}
    \centering
    \includegraphics[width=\textwidth]{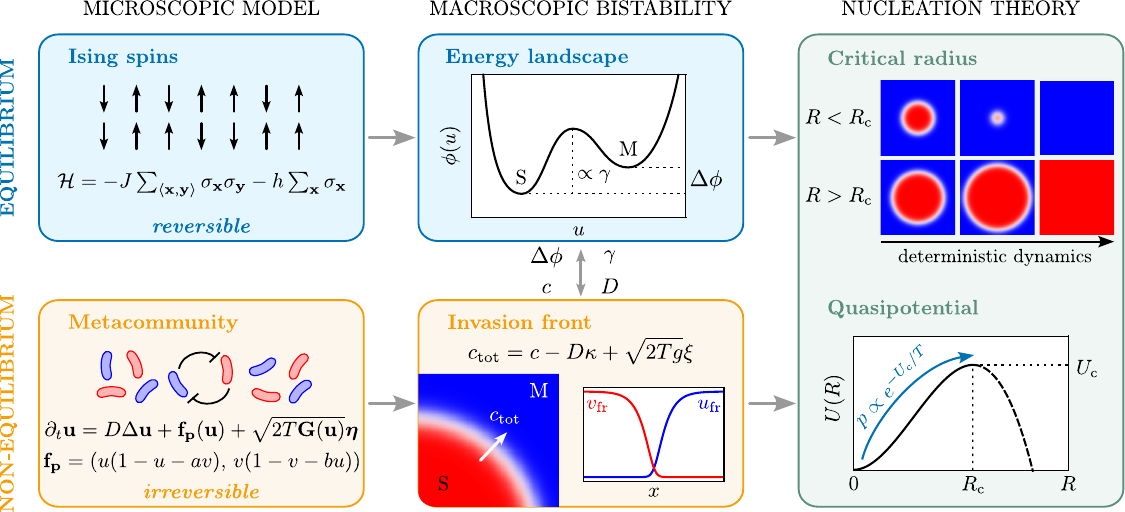}
    \caption{Nucleation in equilibrium and non-equilibrium systems. Microscopic models, whether equilibrium (blue), like the Ising model, or non-equilibrium (yellow), like ecological metacommunities, share common features: bistability due to local dynamics and spatial coupling. In equilibrium, the energy density difference $\Delta\phi$ and the surface tension $\gamma$ determine the nucleation properties: the critical radius $R_\mrm c$ above which droplets grow deterministically, and the probability that noise drives subcritical droplets to grow beyond $R_\mrm c$. Out of equilibrium, where no energy landscape is available, these quantities are replaced by dynamical properties of propagating fronts. The total speed of the front, $c_\mrm{tot}$, is corrected with respect to its flat speed, $c$, by a term involving its response $D$ to curvature $\kappa$, and a white noise $\xi$ weighted by the mobility $g$. Both $c$ and $g$ are integral properties of the front profile and encode its internal structure. The resulting nucleation theory for non-equilibrium systems has the same mathematical structure as in equilibrium (green): $c$ and $D$ play the roles of $\Delta\phi$ and $\gamma$, respectively. At equilibrium, the front properties are directly related to their energetic counterparts.}
    \label{fig1}
\end{figure*}

The central question of this work is how the nucleation theory valid for equilibrium systems extends to a family of non-equilibrium systems without an underlying energy landscape. We consider a general class of stochastic reaction-diffusion models with a non-conserved vector order parameter, and focus on a bistable Lotka--Volterra model as a concrete ecological realization. We first establish the general nucleation theory for this class of models, and then show how it predicts the outcome of ecological invasions. Figure~\ref{fig1} summarizes the framework developed in this article and its correspondence with equilibrium concepts. 

\subsection{Models and Questions}

\subsubsection{Generic Reaction-Diffusion Models}

We consider stochastic reaction-diffusion systems with a non-conserved, vector, order parameter $\mb u(t,\,\mb x)$. The order parameter is a set of local observables that characterize the state of the system as a function of space $\mb x\in \mathbb{R}^d$ and time $t$. Its dynamical evolution is of the form
\begin{equation}
    \frac{\partial\mb{u}}{\partial t}=D\Delta\mb u+\mb f_\mb p\qty(\mb u)+\sqrt{2T\mb{G}\qty(\mb u)}\boldsymbol{\eta}\qty(t,\,\mb x)\text.\label{eq:rd}
\end{equation}

The $d$-dimensional Laplacian $\Delta$ describes spatial coupling through diffusion, with a coefficient $D$. The local deterministic dynamics are described by the force field $\mb f_\mb p$, which depends on a set of parameters $\mb p$. We assume that, for all parameter values considered, $\mb f_\mb p$ has exactly two stable fixed points, $\mb u_+^*$ and $\mb u_-^*$, corresponding to the two alternative stable states. We call binodal the set of values of $\mb p$ such that $\mb u_-^*$ and $\mb u_+^*$ are equally stable, in a sense that will be made precise in Section~\ref{sec:frontsgeneral}~\cite{pomeau_front_1986}. Off the binodal, we refer to $\mb u_+^*$ as the metastable state. The Gaussian white noise $\boldsymbol{\eta}$ has correlations
\begin{equation}
    \ev{\eta_i\qty(t,\,\mb x)\eta_j\qty(t',\,\mb x')}=\delta_{ij}\delta\qty(t-t')\delta^d\qty(\mb x-\mb x')\text,
    \label{eq:noisecorr}
\end{equation}
and the matrix $\mb G(\mb u)$ is symmetric and positive, such that its square root is unambiguously defined. The overall noise amplitude is quantified by the parameter $T$. We work in the weak-noise limit $T\to0$, where Stratonovich and Itō conventions are equivalent~\cite{gardiner_handbook_1983}. The extension to finite noise strength, and the challenges it poses, are discussed in \ref{sec:limits}.

This general class of models encompasses both equilibrium and non-equilibrium dynamics. In equilibrium, the deterministic force derives from a potential and the noise and mobility satisfy the appropriate fluctuation-dissipation relation. We impose neither condition here: the local dynamics $\mb f_\mb p$ need not derive from a potential, and the noise amplitude $\mb G$ may depend on the state $\mb u$. Our framework therefore applies to systems that break detailed balance, including ecological models in which local demographic dynamics (the reaction term) are coupled across space by dispersal (diffusion) and subjected to demographic or environmental fluctuations (the noise).

\subsubsection{The Lotka--Volterra Model}

As a concrete ecological realization of Eq.~(\ref{eq:rd}), we consider the archetypal two-species Lotka--Volterra model for spatial competition between two antagonistic species~\cite{goel_volterra_1971}. The order parameter, $\mb u=\mqty(u&v)^\mathsf T$, quantifies the abundances per unit space of each species, $u$ and $v$. We use rescaled abundances and time units such that the carrying capacities and intrinsic growth rates are unity. We consider the LV population dynamics
\begin{equation}
    \mb f_\mb p\qty(\mb u)=\mqty(u\qty(1-u-av)\\v\qty(1-v-bu))\text,\label{eq:lv}
\end{equation}
which combines a logistic growth term and local, quadratic, interactions characterized by strength coefficients $\mb p=\mqty(a&b)^\mathsf T$. These effective parameters may encompass different biological mechanisms, including competition for resources or antagonism through the emission of toxins~\cite{hibbing2010bacterial,dean_2019}. The matrix $\mb G$ may encode fluctuations of demographic or environmental origins, as discussed in Sec.~\ref{sec:applicationlv}. 

The LV system is bistable for $a>1$ and $b>1$, and the two stable states correspond to the exclusive presence of either species, 
\begin{equation}
    \mb u_+^*=\mqty(1\\0)\text, \qquad \mb u_-^*=\mqty(0\\1)
\end{equation}
when $a>b$, see Sec.~\ref{sec:portraitlv}.

We remain agnostic about the microscopic processes that determine the carrying capacities and interaction coefficients. Our purpose in introducing the LV model is therefore not to provide a fully general or realistic description of ecological communities, but rather a minimal and tractable model in which the effects of local antagonistic interactions on invasion can be isolated and interpreted. 

A commonly used approximation reduces the full vector description to a single scalar order parameter by assuming that the total abundance $u+v$ remains constant. The resulting variable, the frequency $f=u/\qty(u+v)$, then obeys dynamics equivalent to that of a single species with a strong Allee effect~\cite{lavrentovich_asymmetric_2014, tanaka_spatial_2017, lavrentovich_nucleation_2019}. We do not make this approximation, because $u+v$ in fact varies across an invasion front, and this variation directly shapes the front dynamics and the nucleation probability.

\subsubsection{Main Questions}

Our goal is twofold. First, we seek a general theory of homogeneous nucleation for the broad class of systems described by Eq.~(\ref{eq:rd}). Working in the weak-noise limit, we ask for the critical radius $R_c$ that separates deterministic growth from decay and, for a sub-critical droplet of the stable state ($R_0<R_c$), for the probability $p(R_0)$ that noise nonetheless drives it to grow and invade the surrounding metastable state. Second, we investigate how local ecological interactions shape the invasion of one species into a community dominated by the other: specifically, how the competitive advantage of one species and the strength of their antagonistic interactions set the front dynamics, the critical inoculum size, and the probability of successful invasion.
 
\subsection{Nucleation Theory Beyond Equilibrium}

Our general theoretical conclusion is that the mathematical structure of Classical Nucleation Theory (CNT) extends well beyond equilibrium. For a broad class of reaction-diffusion systems without conserved dynamics, the standard CNT equations remain valid provided the thermodynamic quantities are replaced by dynamical properties of propagating fronts. These fronts are nonlinear solutions describing the spatial variation of the order parameter $\mb u$ across the interface separating the metastable and stable states. We illustrate in Fig.~\ref{fig1} the similarities and differences between nucleation theory in and out of equilibrium in the ``classical'' regime.

Specifically, the speed $c$ of a flat front determines which state invades the other and therefore quantifies metastability, playing the role of the equilibrium free-energy difference $\Delta\phi$. Likewise, the diffusion coefficient $D$ couples neighboring regions along the interface and thus governs its response to curvature, in direct analogy with the equilibrium surface tension $\gamma$. 
These correspondences carry through the whole structure of CNT, yielding the expressions for the activation barrier, the critical radius and the invasion probability summarized in Table~\ref{tab:correspondence}. Extending CNT out of equilibrium thus amounts largely to identifying the dynamical quantities that take the place of the usual thermodynamic variables.

\begin{table}
    \centering
    \begin{tabular}{cc}
        \textbf{Equilibrium}&\textbf{Non-equilibrium}\\[3pt]
        \hline
        &\\[-2pt]
        $\Delta\phi$&$c$\\[3pt]
        $\gamma$&$D$\\[3pt]
        $\displaystyle p\propto\exp\qty(-\frac{E}{T})$&$\displaystyle p\propto\exp\qty(-\frac{U}{T})$\\[10pt]
        $\displaystyle R_\mrm c=\qty(d-1)\frac{\gamma}{\Delta\phi}$&$\displaystyle R_\mrm c=\qty(d-1)\frac{D}{c}$\\[10pt]
        $\displaystyle E=S_dR^{d-1}\qty(\gamma-\frac{\Delta\phi}{d}R)$\makebox[5pt]{}&\makebox[5pt]{}$\displaystyle U=\frac{1}{g}S_dR^{d-1}\qty(D-\frac{c}{d}R)$
    \end{tabular}
    \caption{Correspondence between classical nucleation theory at equilibrium (left) and its extension to non-equilibrium systems with non-conserved order-parameter dynamics (right), derived in this article. The mathematical structure of CNT is preserved: equilibrium energetic quantities are replaced by dynamical properties of propagating fronts. The front speed $c$ plays the conceptual role of the free-energy difference $\Delta\phi$, while the front diffusivity $D$ plays that of the surface tension $\gamma$. This dictionary extends to the expressions for the invasion probability $p$, critical nucleus radius $R_c$ and quasipotential $U$.}
    \label{tab:correspondence}
\end{table}

The key step underlying these results is a description of propagating fronts close to the binodal and for weak curvature. By quantifying their response to curvature $\kappa$ and fluctuations, we coarse-grain away their internal structure and derive an effective interface dynamics. Remarkably, the interface dynamics takes the same form irrespective of whether the underlying microscopic dynamics satisfies detailed balance. The resulting effective description is summarized in Fig.~\ref{fig1}: the speed of the curved, noisy interface
\begin{equation}
    c_\mrm{tot}=c-D\kappa+\sqrt{2Tg}\xi\text,
\end{equation}
is governed by the speed $c$ of a flat front, its response $D$ to curvature $\kappa$, and fluctuations comprising a gaussian white noise $\xi$, uncorrelated on scales larger than the front width, weighted by $g$. The coefficients $c$ and $g$ can be expressed directly in terms of the front profile, providing a bridge between the microscopic dynamics and the effective nucleation theory. 

This framework applies, as in ``classical'' nucleation, in the regime where the speed of the interface $c$ is small and the radius of the critical nucleus is much larger than the interface width. This separation of scales disappears as the spinodal is approached, \textit{i.e.} when the metastable state $\mb u_+^*$ is on the verge of stability. We show that spinodal nucleation theory can nevertheless be generalized out of equilibrium. The critical nucleus is obtained perturbatively from the metastable state $\mb u_+^*$ and aligns with the marginal (least stable) eigenmode of the Jacobian of $\mb f_\mb p$, effectively reducing the dynamics to that of a single order parameter.

We validate these theoretical predictions through extensive numerical simulations of the Lotka--Volterra model. Beyond providing a stringent test of our theory, this application illustrates how the front-based framework connects nucleation properties to experimentally accessible ecological parameters.

\subsection{Application to a Bistable Antagonistic Ecological System}

In the Lotka--Volterra model, nucleation becomes synonymous with invasion: suppose that an inoculum of one species is introduced into an environment dominated by another species. Even if the invader is a stronger competitor at equal abundances, invasion may fail if the inoculum is too small due to the overwhelming presence of the resident competitor.

We find that the speed of fronts, the critical radius of inocula and the probability of invasion result from a tradeoff between the competitive advantage of the better competitor, $a-b$, which acts as a driving force for fronts, and their susceptibility to deterministic and stochastic perturbations, which is directly linked to their inner structure and controlled by $a+b$.

In particular, in strongly competing communities characterized by large values of $a+b$, the front develops a pronounced depletion region. Within this region, the two species coexist only over short length scales, and the total abundance $u+v$ is substantially lower than the carrying capacity. Because this region is the locus of interspecific competition, its structure strongly influences the speed of invasion. Consequently, for a given competitive advantage $a-b$, stronger interactions lead to slower fronts, larger critical inocula and lower invasion probabilities. 

This dependence can be made quantitative. In two dimensions, the probability that a small inoculum, of radius comparable to the front width, successfully invades scales approximately as $\ln p \propto -(a+b-2)/(a-b)$, so that at fixed competitive advantage $a-b$, invasion is suppressed exponentially as the competition strength $a+b$ increases (Sec.~\ref{sec:applicationlv}). 

These results show that the LV model is more than a natural application for our theory: the two species' abundances form a genuinely vectorial order parameter. We demonstrate that the vectorial internal structure of the front carries information essential for capturing the ecological consequences of strong antagonism. Such information would be lost upon reduction to a single scalar order parameter $f$. 

Our results therefore establish a direct link between local ecological interactions and macroscopic nucleation properties through the structure and dynamics of propagating fronts. 

\section{Nucleation theory: from equilibrium to non-equilibrium\label{sec:nuceq}}

In this section, we develop the route toward a nucleation theory for the general class of non-equilibrium models described by Eq.~(\ref{eq:rd}). We begin with the equilibrium special case in Sec.~\ref{subsec:equilibriumnucleation}. This part is standard and can be found for instance in Ref.~\cite{debenedetti_metastable_2020}. We use it as a pedagogical reference point for the non-equilibrium theory that follows. We recall how detailed balance allows the probability of a rare transition to be reduced to an energy-barrier crossing problem, yielding the familiar scalings of classical nucleation theory. This reduction is precisely what fails out of equilibrium: without detailed balance, there is no energy landscape from which to read transition probabilities. We therefore turn in Sec.~\ref{sec:largedev} to large-deviation theory, which provides the natural framework for describing the probability of rare transitions and for identifying the dynamical quantities that replace energetic ones.

\subsection{Nucleation Theory at Equilibrium}\label{subsec:equilibriumnucleation}

A first-principles theory of nucleation was derived independently in two areas of physics: by Langer, to describe the condensation of supersaturated vapors and, more generally, the decay of metastable states~\cite{langer_theory_1967, langer_statistical_1969}, and by Coleman and Callan, to describe the decay of the false vacuum in high-energy physics~\cite{coleman_aspects_1985}. These field-theoretic derivations were later complemented by the rigorous mathematical framework developed by Freidlin and Wentzell to study random perturbations of dynamical systems~\cite{freidlin_random_2012}, which underlies the large-deviation approach we use below. 

\subsubsection{The Field Equation}

Let us first consider the special case in which Eq.~(\ref{eq:rd}) describes equilibrium dynamics. We thus write $u$, instead of $\mb u$, the relevant scalar order parameter. The local force derives from a potential, $f_\mb p = -\alpha \phi'(u)$, and the noise strength is tied to the mobility $\alpha$. Equation~(\ref{eq:rd}) then reduces to the Ginzburg--Landau equation
\begin{align}
    \frac{\partial u}{\partial t}&=\alpha\Gamma\Delta u-\alpha\phi'\qty(u)+\sqrt{2\alpha T}\eta\qty(t,\,\mb x)\\
    &=-\alpha\frac{\delta\mathcal E}{\delta u\qty(\mb x)}+\sqrt{2\alpha T}\eta\qty(t,\,\mb x)\text,
    \label{eq:gl}
\end{align}
with
\begin{equation}
    \mathcal E\qty[u\qty(\mb x)]=\int\phi\qty(u\qty(\mb x))+\frac{1}{2}\Gamma\abs{\boldsymbol{\nabla}u}^2\mrm d\mb x\text.
\end{equation}
Here $\phi(u)$ is the free-energy density of a homogeneous phase of order parameter $u$, which takes the form of a double well with minima $u^*_\pm$ when the system is bistable. An example is the ferro--paramagnetic transition in spin systems, shown in Fig.~\ref{fig1}, for which $u$ is the magnetization. The coefficient $\Gamma$ penalizes inhomogeneity. It is often phenomenological but, in some cases, can be derived from microscopic interactions. Equation~(\ref{eq:gl}) is a relaxational gradient descent on $\mathcal E$, corresponding to Model A in the Hohenberg-Halperin classification~\cite{hohenberg_theory_1977}. It can describe many first-order phase transitions whose order parameter is not conserved. The fluctuation-dissipation relation, which ties the noise strength to the mobility, is what makes the dynamics obey detailed balance, with the Boltzmann steady state $p[u]\propto\exp(-\mathcal E[u]/T)$, where the temperature $T$ is expressed in units of energy ($k_\mrm B=1$).

\subsubsection{The Critical Nucleus}

A central object of field theories of nucleation is the \emph{critical nucleus} $u_\mrm c(\mb x)$~\cite{cahn_free_1959, bates_dynamics_1993, muratov_threshold_2013, b_muratov_threshold_2017}. It resembles an extended droplet, and corresponds to a stationary configuration of the deterministic dynamics. As such, the critical nucleus does not depend on the noise. 

The critical nucleus admits two equivalent characterizations. First, it is a fixed point of Eq.~(\ref{eq:gl}) with exactly one unstable mode: perturbations on one side of the unstable direction lead to decay toward the metastable state, while perturbations on the other lead to growth toward the stable state. The critical nucleus thus results from a balance between effects that tend to favor nucleation, associated with the local term $\phi$, and those that tend to hinder it, captured by the gradient term $\Gamma$. Second, it is the configuration that minimizes the energy barrier between the metastable and stable states. The critical nucleus is therefore the field configuration through which nucleation is most likely to occur.

Importantly, the critical nucleus $u_\mrm c(\mb x)$ is a full field configuration, which cannot \textit{a priori} be reduced to its scalar critical radius. Its full structure will therefore be essential to the nucleation theory developed below. Because Eq.~(\ref{eq:rd}) is translationally invariant, the critical nucleus is defined only up to translation; we define $\mb x=\mb 0$ to be its center.

\subsubsection{The Nucleation Probability}

The critical nucleus separates deterministic decay from deterministic growth. Thus, reaching it from a subcritical droplet requires a fluctuation against the deterministic dynamics. At equilibrium, detailed balance provides a remarkable simplification: it relates the probability of a fluctuation-driven trajectory to that of its time reverse, which follows the deterministic dynamics. As a consequence, the probability of reaching the critical nucleus can be calculated using only the energy difference between the initial subcritical configuration and the critical nucleus, and does not require knowing the typical trajectory taken~\cite{bouchet_generalisation_2016}.

More precisely, detailed balance relates the probability $p_{\Delta t}[u]$ of observing a particular evolution $u\qty(t,\,\mb x)$ of the system from $t=0$ to $t=\Delta t$ to the probability $p_{\Delta t}[\overline{u}]$ of observing its time reverse $\overline{u}(t):=u\qty(\Delta t-t)$, namely
\begin{equation}
    p_{\Delta t}[u] = p_{\Delta t}[\overline{u}]\,e^{\displaystyle -\frac{\mathcal E\qty[u\qty(\Delta t)]-\mathcal E\qty[u\qty(0)]}{T}}\text.\label{eq:db}
\end{equation}

Consider now a trajectory $u\qty(t,\,\mb x)$ that starts from a subcritical configuration $u(0)=u_0$ and reaches the critical nucleus, $u(\Delta t)=u_c$. This forward trajectory describes the fluctuation-driven growth of the droplet against its deterministic tendency to shrink. Its time reverse, $\overline{u}\qty(t,\,\mb x)$, instead follows the deterministic dynamics from the critical nucleus toward the metastable state and therefore has high probability. Since the exponent in Eq.~(\ref{eq:db}) does not depend on the path taken, the probability $p\qty(u_0)$ that a droplet starting from $u_0$ reaches the critical nucleus is 
\begin{equation}
    p\qty(u_0)\propto\exp\qty(-\frac{\mathcal E\qty[u_\mrm c]-\mathcal E\qty[u_0]}{T})\text,
    \label{eq:arrhenius}
\end{equation}
which is the Arrhenius law.

At equilibrium, the nucleation problem thus reduces to two pieces of information: the critical nucleus $u_\mrm c$, the configuration which separates decay from growth, and its energy $\mathcal E\qty[u_\mrm c]$, which determines the probability of reaching it. In general, the critical nucleus cannot be determined analytically, but important simplifications arise in appropriate limits. 

\subsubsection{Classical Nucleation Theory}

Close to the binodal, when $\Delta\phi=\phi\qty(u_+^*)-\phi\qty(u_-^*)$ is small, the two homogeneous states are nearly equally stable and the driving force for nucleation is weak. A large critical nucleus is therefore required for the bulk energy gain associated with converting the metastable phase into the stable one to overcome the energetic cost of the interface~\cite{bray_theory_2002}. Its radius $R_\mrm c$ is thus larger than the width $w$ of the interface, $R_\mrm c \gg w$. This separation of length scales allows for the critical nucleus to be approximated as a spherical droplet: its interior is exponentially close to the stable state $u_-^*$, while its interface is locally equivalent to a flat interface between the two phases. This approximation is the basis of classical nucleation theory. 

Let $u_R$ denote a configuration consisting of a spherical droplet of radius $R$ of the stable state inside the metastable state, and $u_+^*$ the homogeneous metastable state. The energy barrier associated with forming the droplet is 
\begin{equation}
\mathcal E\qty[u_R] - \mathcal E\qty[u_+^*] \approx \mathcal E_\mrm{bulk}+\mathcal E_\mrm{interface}\text.
\end{equation}
The bulk gain is proportional to the volume of metastable state replaced by the stable one, $\mathcal E_\mrm{bulk} = - S_d R^d \Delta \phi/d$, with $S_d$ the surface area of the unit sphere in $d$ dimensions. The interfacial contribution is proportional to the surface area, $\mathcal E_\mrm{interface} = S_dR^{d-1}\gamma$, with $\gamma$ the surface tension of a flat interface. It can be expressed in terms of the interfacial profile $u_R\qty(r)$ as ~\cite{cahn_free_1958}
\begin{align}
    \gamma&=\int_{R-w/2}^{R+w/2}\phi\qty(u_R\qty(r))-\phi\qty(u_+^*)+\frac{1}{2}\Gamma\qty(\frac{\partial u_R}{\partial r})^2\mrm dr\\
    &=\int_{R-w/2}^{R+w/2}\Gamma\qty(\frac{\partial u_R}{\partial r})^2\mrm dr\text.\label{eq:gamma}
\end{align}

In the thin-interface limit of CNT, the energy barrier is a function of the droplet radius, $E(R)$, with a shape similar to $U\qty(R)$ plotted in Fig.~\ref{fig1}. It grows until a critical radius $R_\mrm c$ before decreasing, resulting in an activation barrier $E_\mrm c = E(R_\mrm c)$ directly related to $\Delta \phi$ and $\gamma$. The CNT expressions for $E(R)$ and $R_\mrm c$ are provided in Table~\ref{tab:correspondence}. As illustrated in Fig.~\ref{fig1} (right), droplets with initial radius $R<R_\mrm c$ decay deterministically, but those with $R>R_\mrm c$ grow and invade space. 

Yet, rare fluctuations may bring a subcritical droplet of initial radius $R_0 < R_\mrm c$ to the critical nucleus, as depicted by the blue arrow in Fig.~\ref{fig1}. In the CNT limit, computing the nucleation probability $p(R_0)$ is simple: it follows directly from Eq.~(\ref{eq:arrhenius}),
\begin{equation}
p(R_0)\propto
\exp\qty(-\frac{E(R_\mrm c)-E(R_0)}{T})\text.
\label{eq:CNTarrhenius}
\end{equation}
In particular, the probability of spontaneous nucleation ($R_0=0$) is controlled by the activation barrier $E_\mrm c$, as reported in Table~\ref{tab:correspondence}. 

The CNT derivation highlights the central role of the energetic quantities $\Delta\phi$ and $\gamma$. Our generalization beyond equilibrium will seek their dynamical counterparts, while preserving the distinction between bulk driving and interfacial resistance that underpins the theory.

\subsubsection{Spinodal Nucleation Theory}

When $\Delta\phi$ is not small, no systematic simplification of the field equation can be made and the critical nucleus must be determined numerically, although some models can be formulated depending on the problem at hand~\cite{cahn_free_1959, gunton_homogeneous_1999, karthika_review_2016, lutsko_how_2019, wu_nonclassical_2026}. In particular, the critical nucleus cannot be seen as a large droplet with sharp edges. As such, its radius is not enough to characterize it entirely, ruling out the possibility of using CNT. This regime is sometimes known as non-classical nucleation.

However, close to the spinodal limit, defined as the point where the metastable state becomes unstable, the problem of finding the critical nucleus and computing its energy becomes tractable again and makes up a theory known as spinodal nucleation, which we now recall~\cite{cahn_free_1959, unger_nucleation_1984, muratov_breakup_2004}.

Beyond the spinodal, the homogeneous state $u_+^*$ becomes unstable and long-wavelength fluctuations grow spontaneously through continuous ordering. Near the spinodal transition, we expect the critical nucleus to resemble these incipient fluctuations, \textit{i.e.} its radius grows large while its amplitude shrinks, such that the energy barrier vanishes at the spinodal. This motivates an expansion around the unstable homogeneous state, $u_\mrm c=u_+^*+\delta u$, so the bulk energy density is
\begin{equation}
    \phi\qty(u_\mrm c)\simeq \phi\qty(u_+^*)+\frac{1}{3!}\phi^{(3)}\qty(u_+^*)\delta u^3+o(\delta u^4) \text,
\end{equation}
where  the constant $\phi\qty(u_+^*)$ is irrelevant, the first derivative is zero since $u_+^*$ is a minimum of $\phi$, and the second derivative vanishes at the spinodal. The expansion must be truncated at the first nonzero order $\delta u^n$, generically $n=3$, but possibly a higher power due to symmetries or fine-tuning, as near a tricritical point.

After appropriate rescalings of space and $\delta u$, this yields the critical nucleus and its universal shape. Systems with the same value of $n$ therefore share the same asymptotic nucleus shape near the spinodal, independently of the detailed form of $\phi$ away from $u_+^*$.

This universality actually only holds below a critical dimension $d_\mrm c(n)$~\cite{muratov_breakup_2004}, in particular $d_\mrm c(3)=6$. Above, the universal description breaks down and the details of $\phi$ become relevant again. We clarify the fate of this universality in systems with vector order parameters in Sec.~\ref{sec:noneqspin}.

\subsection{Towards Non-Equilibrium Nucleation Theory}\label{sec:largedev}

Although the diffusive and bistable local terms arise from different mechanisms in physical and ecological systems, the similarity between Eqs.~(\ref{eq:rd}) and (\ref{eq:gl}) suggests that a common theory may describe nucleation in both settings. At equilibrium, detailed balance underlies the derivation at every step: the rare transition to the critical nucleus $u_c$ is controlled by its energy $E_c$, which determines the probability of reaching it. Out of equilibrium, Eq.~(\ref{eq:rd}) generically violates detailed balance, and no energy landscape is available to characterize transitions~\cite{freidlin_random_2012}. 

New tools are therefore needed to compute the probability that Eq.~(\ref{eq:rd}) evolves from the homogeneous metastable state to the homogeneous stable state. Large-deviation theory provides the natural framework for this problem. It quantifies the probability of noise-induced rare trajectories and leads to exponential scalings analogous to Eq.~(\ref{eq:arrhenius}), with the energy replaced by a dynamical rate function tailored to the dynamics~\cite{touchette_large_2009}.

\subsubsection{Freidlin--Wentzell Action and the Quasipotential}

For diffusion processes such as Eq.~(\ref{eq:rd}), this rate function is obtained by Freidlin–Wentzell theory, which quantifies the probability that the trajectory $\mb u(t)$ deviates from the deterministic one under weak noise~\cite{freidlin_random_2012}. Related rare-event problems have been studied in statistical physics using field-theoretic approaches to master and Langevin equations~\cite{dykman_large_1994, elgart_rare_2004}. These formulations are essentially equivalent. We use the Freidlin--Wentzell framework for its directness and mathematical foundation. 

Concretely, consider the stochastic dynamics
\begin{equation}
    \dot{\mb u}=\mb b\qty(\mb u)+\sqrt{2T\mb G\qty(\mb u)}\eta\qty(t)\text,
\end{equation}
where the dot denotes a time derivative, and $\mb u$ may represent a finite-dimensional vector or, as in Eq.~(\ref{eq:rd}), a field. The probability of observing a trajectory $\mb u(t)$ scales as
\begin{equation}
    p[\mb u]\propto\exp\qty(-\mathcal A\qty[\mb u\qty(t)]/T)\text,
    \label{eq:ratewf}
\end{equation}
where
\begin{equation}
    \mathcal A\qty[\mb u\qty(t)]=\frac{1}{4}\int\abs{\mb G\qty(\mb u)^{-\frac{1}{2}}\qty{\dot{\mb u}-\mb b\qty(\mb u)}}^2\mrm dt\text.
\end{equation}
is the Freidlin--Wentzell action for the trajectory \footnote{In the mathematical literature, the Freidlin--Wentzell action is often defined with a factor $1/2$ rather than $1/4$. Our convention ensures that $U_{\mb u^*}$ coincides with $E$ for equilibrium dynamics close to $\mb u^*$.}. It quantifies a cost associated to noise driving the trajectory away from its deterministic dynamics: the further $\dot{\mb u}$ is from $\mb b(\mb u)$, the less probable the trajectory. Figure \ref{fig2} (a) illustrates a noisy trajectory fluctuating around the deterministic one when $\mb u$ is two-dimensional. 

Violation of detailed balance implies that to compute the transition probability from a fixed point $\mb u^*$ to any configuration $\mb u_0$, one must consider all possible trajectories between the two states. At weak noise, the exponential scaling in Eq.~(\ref{eq:ratewf}) allows a major simplification to occur: among all of them, the transition probability is dominated by the trajectory with the smallest Freidlin--Wentzell action, known as the Minimum Action Path (MAP). The corresponding minimum action defines the quasipotential
\begin{equation}
    U_{\mb u^*}\qty(\mb u_0)=\min_{\substack{\mb u\qty(t),\\\mb u^*\to\mb u_0}}\mathcal A\qty[\mb u\qty(t)]\text.
\end{equation}
Thus, the quasipotential is the dynamical analogue of the energy difference in Eq.~(\ref{eq:arrhenius}): it determines the leading exponential scaling of the probability of reaching $\mb u_0$, while the MAP identifies the trajectory along which this transition is most likely to occur.

We have defined the quasipotential locally, relative to a starting state $\mb u^*$; a global quasipotential, related to the steady-state distribution, can also be defined, and associated to a Hamilton–Jacobi equation~\cite{freidlin_random_2012}. We use only two operational characterizations of $U$: as a transition probability and as the action of the MAP.

The Freidlin--Wentzell theory is only valid in the weak noise limit. We discuss consequences of finite noise levels in Sec. \ref{sec:limits}.

\subsubsection{Nucleation as an Escape Problem}

\begin{figure}
    \centering
    \includegraphics[width=\linewidth]{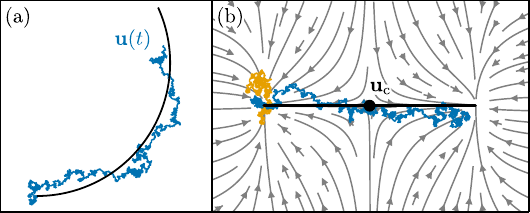}
    \caption{Illustration of Freidlin--Wentzell theory. (a) Deterministic trajectory and a noisy trajectory. (b) Trajectories remain near an attractor for long times (orange) until, on a rare occasion, the noise conspires to push the system across to the other attractor (blue); these unlikely trajectories concentrate around the MAP (black). A general feature of escape problems, which survives in the case of many degrees of freedom, is that the MAP crosses the saddle point separating the two basins. For Eq.~(\ref{eq:rd}) that saddle is precisely the critical nucleus $\mb u_\mrm c$, which therefore retains the role it plays at equilibrium: the most likely configuration to cross the separatrix between basins of attraction.}
    \label{fig2}
\end{figure}

Nucleation can be viewed as an escape problem~\cite{hanggi_1990}: the system described by Eq.~(\ref{eq:rd}) must leave the basin of attraction of the homogeneous metastable state $\mb u_+^*$ and reach the basin of the homogeneous stable state $\mb u_-^*$. In the weak-noise limit, the transition is dominated by a MAP that crosses the saddle separating these two basins (Fig.~\ref{fig2}(b)). For Eq.~(\ref{eq:rd}), this saddle is the critical nucleus $\mb u_\mrm c$ which still plays a central role out of equilibrium as the configuration separating deterministic decay from deterministic growth. However, unlike at equilibrium, knowing $\mb u_\mrm c$ alone is no longer sufficient to determine the probability of reaching it.

The relevant transition is thus the one from the homogeneous metastable state $\mb u_+^*$ to the critical nucleus $\mb u_\mrm c$. The corresponding quasipotential is
$U_\mrm c = U_{\mb u_+^*}\qty(\mb u_\mrm c)$, and the nucleation probability is given by the Freidlin--Wentzell theory~\cite{bouchet_generalisation_2016},
\begin{equation}
    \fbox{$\displaystyle p\propto\exp\qty(-U_\mrm c/T)$}\text,
    \label{eq:defuc}
\end{equation}
and reported in Table~\ref{tab:correspondence}.

The central object of non-equilibrium nucleation theory is therefore not a static configuration and its energy barrier, but the MAP connecting the metastable state to the critical nucleus and its quasipotential. This makes non-equilibrium nucleation theory challenging: the MAP is a path through the infinite-dimensional space of field configurations. At equilibrium this dimensionality problem can be already substantial, since one needs to find a static field configuration saddle $\mb u_ c$; out of equilibrium it can be daunting, because the quasipotential is obtained by a minimization over entire trajectories in function space.

\subsubsection{Nucleation Theory in Two Limiting Regimes}

To make this problem tractable, we exploit simplifications that arise in two limiting regimes, inspired by CNT and spinodal nucleation theory. 

Close to the binodal, we expect bulk deviations from $\mb u^*_\pm$ to dominate the action, since the local force field $\mb f_\mb p$ acts throughout the system to suppress them. As in CNT, the MAP should thus to consist primarily of homogeneous regions at $\mb u^*_\pm$, separated by thin interfaces. If, as suggested by the structure of the critical nucleus, most of the MAP involves weakly curved interfaces, their profiles should remain close to those of flat fronts. Our approach to non-equilibrium nucleation theory close to the binodal is therefore to coarse-grain the field $\mb u\qty(\mb x)$ and retain only the interface positions as dynamical variables, as we develop in Sec.~\ref{sec:noneqbin}.

Close to the spinodal, in contrast, the critical nucleus becomes small and the separation between bulk and interfacial length scales disappears. The nucleation problem is then controlled by the least stable mode of the metastable state rather than by the dynamics of a well-defined interface. We use this complementary approach to develop a non-equilibrium spinodal nucleation theory in Sec.~\ref{sec:noneqspin}.

\section{Non-Equilibrium Nucleation Theory\label{sec:noneqnuc}}

We now pursue two complementary routes toward a nucleation theory for the non-equilibrium models described by Eq.~(\ref{eq:rd}). Close to the binodal (Sec.~\ref{sec:noneqbin}), the separation between bulk and interfacial length scales allows us to coarse-grain the dynamics onto that of propagating fronts and compute the quasipotential, with CNT recovered when detailed balance holds. Close to the spinodal (Sec.~\ref{sec:noneqspin}), that scale separation is lost, and we generalize the equilibrium spinodal nucleation theory step by step. In Sec.~\ref{sec:limits}, we discuss corrections to our theory, including the nucleation probability prefactor and extensions of our framework.

\subsection{Close to the Binodal\label{sec:noneqbin}}

In this section, we first establish the general properties of flat fronts in the regime close to the binodal: their existence, uniqueness, stability, and shape. We then perturb flat fronts to derive the stochastic dynamics of curved interfaces, showing how curvature and noise enter their motion. This front dynamics provides the effective description needed to compute the quasipotential governing nucleation. Finally, we show that, in the equilibrium limit, our construction reduces to classical nucleation theory.

\subsubsection{Front Properties\label{sec:frontsgeneral}}

Fronts are defined as solutions $\mb u_\mrm{fr}$ of Eq.~(\ref{eq:rd}) that: i) connect spatially the two stable states $\mb u_\pm^*$, ii) depend only on one spatial coordinate $x$, iii) move at constant speed $c$, without deforming. 
Since this implies $\partial\mb u_\mrm{fr}/\partial t=-c\;\mrm d\mb u_\mrm{fr}/\mrm dx$, they solve
\begin{equation}
    -c\dv{\mb{u}_\mrm{fr}}{x}=D\dv[2]{\mb{u}_\mrm{fr}}{x}+\mb f_\mb p\qty(\mb u_\mrm{fr}),\quad\mb u_\mrm{fr}\qty(\pm\infty)=\mb u_\pm^*\text.\label{eq:front}
\end{equation}

Fronts can be viewed as diffuse interfaces between two homogeneous regions, whose profile and speed are selected by the balance between diffusion and local reaction terms.

\textit{Existence.\,---} Front existence is a well-studied mathematical problem for scalar reaction-diffusion equations~\cite{mckean_nagumos_1970, murray_mathematical_1989, cross_pattern_1993}. For vector equations where $\mb f_\mb p$ is not a gradient, standard tools such as the comparison principle fail, so existence must be proven case by case~\cite{fife_asymptotic_1977, tang_propagating_1980, gardner_existence_1982, hosono_singular_1982, gardner_application_1984, mimura_3-component_1986, kan-on_parameter_1995, kan-on_existence_1996, guo_sign_2013}. We take the existence of vectorial fronts for granted, as supported by numerical simulations of the LV model. 

The existence of fronts provides dynamical objects that serve as pivots to study the nonlinear evolution of the whole field, much like fixed points. Solving them analytically is challenging as one should deal with the full nonlinearity of $\mb f_\mb p$. Yet, this is what makes fronts powerful objects: they encapsulate the coupling between that nonlinearity and the spatial dynamics. They are, however, readily obtained by direct numerical simulation (see Appendix~\ref{sec:numfronts}) and can be studied experimentally. 

We thus treat fronts as the relevant dynamical objects to build our theory. We now turn to the front properties that are needed to compute their response to curvature and fluctuations.

\textit{Uniqueness, Stability and Gap.\,---} In scalar bistable systems the front is unique, linearly stable and gapped. It has a well-defined speed $c$ which depends on the field parameters $\mb p$~\cite{mckean_nagumos_1970}. The sign of $c$ determines which homogeneous state invades the other: it may serve as a dynamical definition of metastability even out of equilibrium~\cite{pomeau_front_1986}.
Linear stability implies that small perturbations of the front profile decay back to the same traveling-wave solution in the comoving frame. By gapped, we mean that relaxation occurs at a finite rate, so that the translational mode is isolated from the rest of the spectrum. The front thus behaves as an attractor for the dynamics in the comoving frame, to which convergence is exponentially fast.

This situation contrasts with the monostable case, such as the Fisher--KPP model for ecological range expansions~\cite{kan-on_fisher_1997, vansaarloos_front_2003}. There, fronts exist for a continuum of speeds, and compactly-supported initial conditions select the slowest one~\cite{lucia_linear_2004}. Such fronts are marginally stable and convergence to them is only algebraic~\cite{gallay_local_1994, faye_asymptotic_2019}.

We assume that fronts in vectorial bistable systems are unique, linearly stable and gapped, as in the scalar bistable case. Numerical study of fronts in the LV model supports this assumption. 

\textit{Front profile.\,---} Scalar fronts are generally monotonic in order to be stable~\cite{murray_mathematical_1989}, unless higher-order spatial derivatives are considered~\cite{vansaarloos_front_2003}. This strongly restricts the possible profiles they can display. Vector order parameters lift this restriction. We demonstrate that the inner structure of vectorial fronts has major implications for their dynamics, highlighting the importance of considering vector order parameters to preserve complex front profiles.

\subsubsection{Stochastic Motion of Curved Fronts}

\begin{figure}
    \centering
    \includegraphics[width=0.8\linewidth]{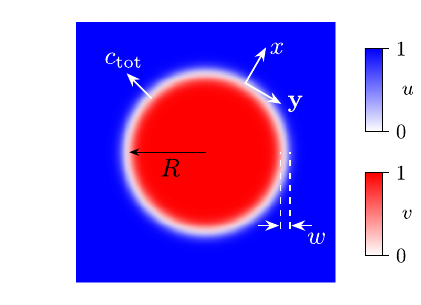}
    \caption{Nucleating droplet in the Lotka--Volterra model. An inoculum of radius $R$ of the stable state $v=1$ (red) in a metastable background $u=1$ (blue). The interface, of width $w$, propagates at speed $c_{tot}$ along direction $x$. Directions orthogonal to propagation are noted $\mb y$. Close to the binodal line, $R \gg w$.}
    \label{fig3}
\end{figure}

Close to the binodal, nucleation proceeds through the growth of a droplet, so the relevant interface is necessarily curved on the scale of the whole nucleus. Yet, we have justified that the critical radius is much larger than the front width, so the interface is locally weakly curved and each local patch can be treated as a weakly-perturbed flat front. Since Eq.~(\ref{eq:rd}) is stochastic, we must also project fluctuations onto the interface motion. Our goal is therefore to derive an effective equation for the stochastic motion of curved fronts.

\textit{Linearization around the flat front solution.\,---} We follow the method of Ref.~\cite{fife_dynamics_1988} and introduce the notations in Fig.~\ref{fig3}. At each point along the interface, we use $x$ for the coordinate, normal to the front, along which it propagates, and $\mb y$ for the $d-1$ coordinates normal to it. Our goal is to find the total normal speed of the perturbed front, which we denote by $c_\mrm{tot}$. 

We consider parameters $\mb p$ such that the system is close to the binodal. We write $\mb p_\mrm b$ the parameters on the binodal, and work in the limit $\varepsilon=\abs{\mb p-\mb p_\mrm b}\to0$. We search for weakly perturbed solutions of Eq.~(\ref{eq:rd}), that we decompose as $\mb u=\mb u_\mrm{fr}^\mrm b+\boldsymbol{\delta}\mb u$, where $\mb u_\mrm{fr}^\mrm b$ is the flat front solution corresponding to $\mb p_\mrm b$, and $\boldsymbol{\delta}\mb u$ is of order $\varepsilon$. In the frame of reference of the front, the time derivative in Eq.~(\ref{eq:rd}) becomes
\begin{equation}
    \evaluate{\frac{\partial\mb u}{\partial t}}{\mb x}=\evaluate{\frac{\partial\mb u}{\partial t}}{x,\mb y}-c_\mrm{tot}\evaluate{\frac{\partial\mb u}{\partial x}}{t,\mb y}\label{eq:dudt}
\end{equation}
and the Laplacian
\begin{equation}
    \Delta_{\mb x}\mb u=\evaluate{\frac{\partial^2\mb u}{\partial x^2}}{t,\mb y}+\kappa\evaluate{\frac{\partial\mb u}{\partial x}}{t,\mb y}+\Delta_{\mb y}\mb u\text,\label{eq:laplacian}
\end{equation}
where $\kappa=\Delta_{\mb x}x$ is the total curvature of the interface and $\Delta_{\mb y}$ is the Laplacian along the curved interface. We write which quantities are held constant while taking the partial derivatives to stress the difference between frames of reference. We consider weakly curved interfaces so that $\kappa$ scales as $\varepsilon$ and we will show that $c_\mrm{tot}$ is also of order $\varepsilon$. Therefore, Eq.~(\ref{eq:dudt}) and Eq.~(\ref{eq:laplacian}) become, to order $1$ in $\varepsilon$,
\begin{equation}
    \evaluate{\frac{\partial\mb u}{\partial t}}{\mb x}=\frac{\partial\boldsymbol{\delta}\mb u}{\partial t}-c_\mrm{tot}\dv{\mb u_\mrm{fr}^\mrm b}{x}
\end{equation}
and
\begin{equation}
    \Delta_{\mb x}\mb u=\frac{\partial^2\boldsymbol{\delta}\mb u}{\partial x^2}+\dv[2]{\mb u_\mrm{fr}^\mrm b}{x}+\kappa\dv{\mb u_\mrm{fr}^\mrm b}{x}
\end{equation}
since $\Delta_{\mb y}\mb u$ scales as $\varepsilon^2$. Finally, we expand at order $\varepsilon$
\begin{equation}
    \mb f_\mrm p\qty(\mb u)=\mb f_\mrm p\qty(\mb u_\mrm{fr}^\mrm b)+\mb J_{\mb p_\mrm b}\qty(\mb u_\mrm{fr}^\mrm b)\cdot\boldsymbol{\delta}\mb u
\end{equation}
with $\mb J_{\mb p_\mrm b}$ denoting the Jacobian of $\mb f_{\mb p_\mrm b}$. Using Eq.~(\ref{eq:front}), we get
\begin{multline}
    \frac{\partial\boldsymbol{\delta}\mb u}{\partial t}-c_\mrm{tot}\dv{\mb u_\mrm{fr}^\mrm b}{x}=\\
   \!\!\!\! \mathcal L_\mrm b\qty[\boldsymbol{\delta}\mb u]+\Delta\mb f\qty(\mb u_\mrm{fr}^\mrm b)+D\kappa\dv{\mb u_\mrm{fr}^\mrm b}{x}+\sqrt{2T\mb{G}\qty(\mb u_\mrm{fr}^\mrm b)}\boldsymbol{\eta}\qty(t,\,\mb x)\text,\label{eq:perturbation}
\end{multline}
where
\begin{equation}
    \Delta\mb f\qty(\mb u_\mrm{fr}^\mrm b)=\mb f_\mb p\qty(\mb u_\mrm{fr}^\mrm b)-\mb f_{\mb p_\mrm b}\qty(\mb u_\mrm{fr}^\mrm b)\label{eq:deltafgeneral}
\end{equation}
and
\begin{equation}
    \mathcal L_\mrm b\qty[\boldsymbol{\delta}\mb u]=D\frac{\partial^2\boldsymbol{\delta}\mb u}{\partial x^2}+\mb J_{\mb p_\mrm b}\qty(\mb u_\mrm{fr}^\mrm b)\cdot\boldsymbol{\delta}\mb u\text.
\end{equation}
The noise $\boldsymbol{\eta}$ has correlations given by Eq.~(\ref{eq:noisecorr}) and is assumed small, so that it may be treated at linear order of perturbation.

In Eq.~(\ref{eq:perturbation}), both $c_\mrm{tot}$ and $\boldsymbol{\delta}\mb u$ are unknowns but only $c_\mrm{tot}$ is of interest. The equation takes the form of a number of source terms, both deterministic like $\Delta\mb f$ and $D\kappa$ as well as the stochastic term, which perturb the front $\mb u_\mrm{fr}^\mrm b$. Its response is dictated by the operator $\mathcal L_\mrm b$, which describes the relaxation of perturbations around it. The term $\Delta\mb f$ involves the distance to the binodal and determines the flat front speed, while the term $D\kappa$ represents the coupling of a front with its neighborhood due to diffusion.

\textit{Solvability condition and front position.\,---} Determining the response of the front to perturbations therefore involves solving an equation of the form
\begin{equation}
    \mathcal L_\mrm b\qty[\boldsymbol{\delta}\mb u]=\text{source terms.}
\end{equation}
However, the operator $\mathcal L_\mrm b$ is singular. Indeed, translational invariance means that any translation of a front is still a solution of Eq.~(\ref{eq:front}), imposing
\begin{equation}
\mathcal L_\mrm b\qty[-\dv{\mb u_\mrm{fr}^\mrm b}{x}]=\mb0\text,
\end{equation}
which can also be checked by direct calculation using Eq.~(\ref{eq:front}).
This reflects the fact that the front is an extended object: it has no intrinsic position, and the definition of its frame of reference is thus ambiguous. Due to this zero mode, perturbations may grow beyond linear order in $\varepsilon$, unless the frame of reference is chosen appropriately.

As in other singular perturbation problems, the appropriate choice for which the singular terms vanish is known as the solvability condition~\cite{bender_advanced_2009, hinch_perturbation_1991}. We choose the scalar product 
\begin{equation}
\left\langle\boldsymbol{\delta}\mb u,\,\boldsymbol{\delta}\mb u'\right\rangle=\int_{-\infty}^{+\infty}\boldsymbol{\delta}\mb u\qty(x)\cdot\boldsymbol{\delta}\mb u'\qty(x)\mrm dx
\text,
\end{equation}
and denote by $\boldsymbol{\delta}\mb u_0$ the eigenvector of the dual of the linearized operator such that $\mathcal L_\mrm b^\dagger\qty[\boldsymbol{\delta}\mb u_0]=\mb0$, defined up to a multiplicative constant. This eigenvector is guaranteed to exist since $\mathcal L_\mrm b^\dagger$ shares the spectrum of $\mathcal L_\mrm b$. The frame of reference of the front is fixed by imposing the solvability condition
\begin{equation}
\left\langle\boldsymbol{\delta}\mb u_0,\,\boldsymbol{\delta}\mb u\right\rangle=0
\text.
\end{equation}

This condition mirrors the classical treatment of a weakly nonlinear oscillator: naive perturbation theory fails because the perturbation shifts the oscillation period~\cite{hinch_perturbation_1991}. This produces resonant terms, singular with respect to the linearized evolution operator, that grow without bound. The fix is the same in both cases: we introduce a shifted quantity (here, the front's frame of reference; there, the oscillator's frequency) as an explicit variable, and choose it so that the resonant terms vanish.

\textit{Stochastic front-motion equation.\,---} We project Eq.~(\ref{eq:perturbation}) on $\boldsymbol{\delta}\mb u_0$ to obtain the stochastic equation for the total front speed
\begin{equation}
    \fbox{$\displaystyle c_\mrm{tot}=c-D\kappa+\sqrt{2Tg}\xi\qty(t,\,\mb y)$}\text,
    \label{eq:frontmotion}
\end{equation}
with\begin{equation}
    c=\frac{\left\langle\boldsymbol{\delta}\mb u_0,\,\Delta\mb f\qty(\mb u_\mrm{fr}^\mrm b)\right\rangle}{\Tilde{\gamma}}\text,
    \label{eq:speed}
\end{equation}
and
\begin{equation}
    g=\frac{\left\langle\boldsymbol{\delta}\mb u_0,\,\mb G\qty(\mb u_\mrm{fr}^\mrm b)\cdot\boldsymbol{\delta}\mb u_0\right\rangle}{\Tilde{\gamma}^2}\text,
    \label{eq:mobility}
\end{equation}
where we introduced 
\begin{equation}
    \Tilde{\gamma}=\left\langle\boldsymbol{\delta}\mb u_0,\,-\dv{\mb u_\mrm{fr}^\mrm b}{x}\right\rangle\text,\label{eq:gammatilde}
\end{equation}
and the term $\xi\qty(t,\,\mb y)$ is a unit gaussian noise with delta correlations in time and no correlations in space on length scales larger than the width of the front.

These equations do not involve $\boldsymbol{\delta}\mb u$: projecting on $\boldsymbol{\delta}\mb u_0$ avoids needing to calculate the deformation of the front and directly extracts the quantities contributing to its speed.

In Eq.~(\ref{eq:speed}), $\boldsymbol{\delta}\mb u_0$ plays the role of a weight against which the driving force $\Delta\mb f$ is integrated across the front. Thus, $\boldsymbol{\delta}\mb u_0$ describes the sensitivity of the front position to perturbations at different points along its profile. The integrated driving force in the numerator of Eq.~(\ref{eq:speed}) acts against the denominator $\Tilde{\gamma}$, which plays the role of a mass for the front in limiting its overall response to forces. The situation is similar for the noise amplitude Eq.~(\ref{eq:mobility}).

Equation (\ref{eq:gammatilde}) reveals that $\Tilde{\gamma}$ is akin to a surface tension in providing an integrated measure of the gradient of $\mb u$ across the front, like $\gamma$ in Eq.~(\ref{eq:gamma}), although it is only defined up to a multiplicative constant.

\textit{Role of the profile deformation.\,---} By projecting Eq.~(\ref{eq:perturbation}) onto $\boldsymbol{\delta}\mb u_0$, we removed dependence on $\boldsymbol{\delta}\mb u$ in the equations. However, we did not neglect $\boldsymbol{\delta}\mb u$, which arises at the same order $\varepsilon$ as $c_\mrm{tot}$. To illustrate the importance of $\boldsymbol{\delta}\mb u$, consider the following formula for $c$ obtained by integrating the flat front equation Eq.~(\ref{eq:front}) against $\mathrm{d} \mb u_\mrm{fr}/\mathrm{d}x$ across the profile of the front,
\begin{equation}
    c=\frac{\int_{\mb u_-^*}^{\mb u_+^*}-\mb f_\mb p\qty(\mb u)\cdot\mrm d\mb u}{\int_{-\infty}^{+\infty}\qty(\dv{\mb u_\mrm{fr}}{x})^2\mrm dx}\text,\label{eq:speedscalar}
\end{equation}
where the integral in the numerator is taken along the front $\mb u_\mrm{fr}\qty(x)$. This equation is routinely used in scalar systems to calculate the speed of a front~\cite{pomeau_front_1986, kado_microscopic_2024}. Now, consider the perturbation of the front by $\Delta\mb f$. For equilibrium dynamics, the numerator would only change by $\int_{\mb u_-^*}^{\mb u_+^*}-\Delta\mb f\qty(\mb u)\cdot\mrm d\mb u$. However, for non-equilibrium dynamics, another term appears due to the deformation of the front because the integral of $\mb f$ depends on the path taken. Eq.~(\ref{eq:speed}), which appears as an alternative to Eq.~(\ref{eq:speedscalar}), brings these two terms together as one thanks to the use of $\boldsymbol{\delta}\mb u_0$ and is therefore better suited for calculations out of equilibrium than Eq.~(\ref{eq:speedscalar}).

\subsubsection{Obtaining the Quasipotential}

\begin{figure}
    \centering
    \includegraphics[width=.95\linewidth]{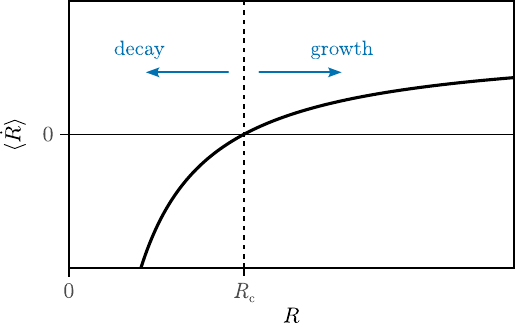}
    \caption{Flow of the droplet radius $R$ according to Eq.~(\ref{eq:motionr}). A critical radius $R_c$ separates deterministic growth from decay. }
    \label{fig4}
\end{figure}

Having coarse-grained the field $\mb u$ and written an equation for the motion of fronts, we are now in position to study the dynamics of spherical droplets involved in nucleation. The evolution of the radius $R$ of the droplet is given at any time by the spatial average of the normal speed $c_\mrm{tot}$ over its surface,
\begin{equation}
    \dot R=\frac{1}{S_dR^{d-1}}\int c_\mrm{tot}\qty(\mb y)\mrm d\mb y\text.
\end{equation}
Since only $\xi$ depends on $\mb y$ in the stochastic front motion Eq.~(\ref{eq:frontmotion}), we get
\begin{equation}
    \dot R=c-D\frac{d-1}{R}+\sqrt{\frac{2Tg}{S_dR^{d-1}}}\zeta\qty(t)\text,
    \label{eq:motionr}
\end{equation}
with $\zeta\qty(t)$ a gaussian noise defined as
\begin{equation}
    \zeta\qty(t)=\frac{1}{\sqrt{S_dR^{d-1}}}\int\xi\qty(t,\,\mb y)\mrm d\mb y\text,
\end{equation}
such that its correlations are $\left\langle\zeta\qty(t)\zeta\qty(t')\right\rangle=\delta\qty(t-t')$. The deterministic part of Eq.~(\ref{eq:motionr}) is plotted in Fig.~\ref{fig4}. A critical radius $R_\mrm c$ separates deterministic growth from decay,
\begin{equation}
    R_\mrm c=\qty(d-1)\frac{D}{c}\text.\label{eq:rc}
\end{equation}

Equation~(\ref{eq:motionr}) can be written as a Langevin equation for the equilibrium dynamics of an effective particle of mobility $\mu\qty(R)$ in a potential $U\qty(R)$ at temperature $T$,
\begin{equation}
    \dot R=-\mu\qty(R)\partial_RU+\sqrt{2T\mu\qty(R)}\zeta\qty(t)\text.
    \label{eq:rdot}
\end{equation}
We identify
\begin{equation}
    \mu\qty(R)=\frac{g}{S_dR^{d-1}}\text,
\end{equation}
and 
\begin{equation}
    \fbox{$\displaystyle U\qty(R)=\frac{S_d}{g}R^{d-1}\qty(D-\frac{c}{d}R)$}\text.\label{eq:potential}
\end{equation}

Equation~(\ref{eq:rdot}) is a one-dimensional Langevin equation, and any such equation satisfies detailed balance with respect to the Gibbs stationary probability distribution, proportional to $\exp\qty(-U/T)$: in one dimension the stationary state carries no probability current. Detailed balance is thus effectively restored once the field has been coarse-grained onto the single variable $R$, and the minimization over trajectories of Sec.~\ref{sec:largedev} goes back, as at equilibrium, to a barrier crossing. What remains of the non-equilibrium nature of Eq.~(\ref{eq:rd}) is contained entirely in the \emph{values} of $c$ and $g$, which no energy function provides. The probability of nucleation from an inoculum of radius $R_0<R_\mrm c$ therefore follows Arrhenius' law
\begin{equation}
    p\qty(R_0)\propto\exp\qty(-\frac{U_\mrm c-U\qty(R_0)}{T})\text,
    \label{eq:arrheniusfronts}
\end{equation}
with
\begin{equation}
    U_\mrm c=U\qty(R_\mrm c)=\frac{\qty(d-1)^{d-1}S_d}{d}\frac{D^d}{gc^{d-1}}\text.\label{eq:uc}
\end{equation}

This completes our goal: we have computed the quasipotential $U_c$ that enters the nucleation probability $p(R_0)$, Eq.~(\ref{eq:defuc}), for the escape of the basin of attraction of the metastable state in Eq.~(\ref{eq:rd}), when $R_0<R_\mrm c$. The corresponding MAP consists in isotropic growth of the droplet, via propagating fronts, until invasion of the entire space. 

\subsubsection{Back to Equilibrium Dynamics: Recovery of CNT\label{sec:eq}}

Once the internal structure of the front has been coarse-grained, its motion takes the same form as for fronts arising from equilibrium dynamics. In particular, the curvature dependence of Eq.~(\ref{eq:frontmotion}) was known since the seminal work of Allen and Cahn on noiseless scalar fronts~\cite{allen_microscopic_1979, coullet_localized_2002}. This does not mean that the front dynamics themselves are at equilibrium: as soon as $c\neq0$, fronts are out of equilibrium. For instance, expanding Eq.~(\ref{eq:frontmotion}) around a flat interface produces a KPZ-like term proportional to $c$. However, details of the field dynamics are entirely absorbed in $c$ and $g$. In a sense, microscopic non-reciprocity is lost when going to the macroscopic scale~\cite{dinelli_non-reciprocity_2023}. 

This matters for non-equilibrium nucleation, because the MAP does not follow the time-reversed relaxation path when the system has to go `uphill' to escape the metastable basin of attraction. However, this lack of time-reversal symmetry is not visible at the level of the front dynamics alone. The relevant difference between the MAP and the relaxation path lies in the small deformations of the internal front profile $\boldsymbol{\delta}\mb u$, which appear jointly, but independently, from the coefficients $c$ and $g$.

Since the front dynamics are the same for equilbrium and non-equilibrium field dynamics, we show that our theory reduces to CNT when Eq.~(\ref{eq:rd}) specializes to equilibrium dynamics. We write $\mb f_\mb p=-\alpha\boldsymbol{\nabla}\phi\qty(\mb u)$ and $D=\alpha\Gamma$, dropping the explicit dependence on $\mb p$ for simplicity and matching the notations of Eq.~(\ref{eq:gl}) with a vector order parameter. Since $\mb f_\mb p$ is a gradient, its Jacobian is a Hessian and the operator $\mathcal L_\mrm b$ thus hermitian, and one may choose $\boldsymbol{\delta}\mb u_0=-\mathrm d\mb u_\mrm{fr}^\mrm b/\mathrm dx$.
The coefficient $\Tilde{\gamma}$ can then be calculated and identified with the interfacial energy $\gamma$ using Eq.~(\ref{eq:gamma}),
\begin{equation}
    \Tilde{\gamma}=\int_{-\infty}^{+\infty}\qty(\dv{\mb u_\mrm{fr}^\mrm b}{x})^2\mrm dx=\frac{\gamma}{\Gamma}\text.
\end{equation}

Using once again $\boldsymbol{\delta}\mb u_0=-\mathrm d\mb u_\mrm{fr}^\mrm b/\mathrm dx$, Eq.~(\ref{eq:speed}) reduces into Eq.~(\ref{eq:speedscalar}) and reads
\begin{equation}
    c=\frac{\int_{-\infty}^{+\infty}-\dv{\mb u_\mrm{fr}^\mrm b}{x}\cdot\Delta\mb f\qty(\mb u_\mrm{fr}^\mrm b)\mrm dx}{\int_{-\infty}^{+\infty}\qty(\dv{\mb u_\mrm{fr}^\mrm b}{x})^2\mrm dx}=\frac{\Gamma}{\gamma}\alpha\int_{\mb u_-^*}^{\mb u_+^*}\boldsymbol{\nabla}\phi\qty(\mb u)\cdot\mrm d\mb u\text,
\end{equation}
yielding
\begin{equation}
    c=D\frac{\phi\qty(\mb u_+^*)-\phi\qty(\mb u_-^*)}{\gamma}:=D\frac{\Delta\phi}{\gamma}\text.
    \label{eq:cdeltaphi}
\end{equation}
Since $\mb G\qty(\mb u)=\alpha\mb 1$, the noise strength, Eq.~(\ref{eq:mobility}), simplifies as
\begin{equation}
    g=\frac{\Gamma^2}{\gamma^2}\int_{-\infty}^{+\infty}\alpha\qty(\dv{\mb u_\mrm{fr}^\mrm b}{x})^2\mrm dx=\frac{D}{\gamma}\text,
\end{equation}
such that the quasipotential Eq.~(\ref{eq:potential}) is equal to the energy barrier,
\begin{equation}
    U\qty(R)=S_dR^{d-1}\qty(\gamma-\frac{\Delta \phi}{d}R)=E\qty(R)\text,
\end{equation}
and the nucleation probability follows the CNT Arrhenius law Eq.~(\ref{eq:CNTarrhenius}).

Thus, the front-based theory preserves the mathematical structure of CNT while replacing equilibrium energetic quantities by dynamical front properties.
The fact that the potential difference $\Delta\phi$ is proportional to the front speed $c$, according to Eq.~(\ref{eq:cdeltaphi}), suggests that $c$ can serve as a proxy for $\Delta\phi$ for non-equilibrium dynamics, where $\Delta\phi$ is no longer available. Like $\Delta\phi$, the front speed $c$ indicates metastability through both its sign and its magnitude. 
The idea that $c$ could play the qualitative role of $\Delta\phi$ out of equilibrium dates back to Pomeau~\cite{pomeau_front_1986}. We showed that this correspondence extends quantitatively to all CNT formulae, as summarized in Table~\ref{tab:correspondence}.

\subsection{Close to the Spinodal\label{sec:noneqspin}}

\subsubsection{Motivation}

We now turn to nucleation close to the spinodal, where the quasipotential can again be obtained analytically, extending arguments previously developed for scalar systems. Before deriving the theory, we first discuss why this regime is of interest.

The spinodal is reached when the metastable fixed point $u_+^*$ approaches linear instability. As this limit is approached, the nucleation energy barrier vanishes. In equilibrium physical phase transitions, however, thermal noise remains finite while the barrier decreases. Consequently, the system typically nucleates before entering the asymptotic spinodal regime, making spinodal nucleation difficult to access experimentally and limiting its practical relevance.

Ecological systems differ in an important respect. Demographic and environmental noise cannot create individuals when a species is absent, so a nearly unstable state may nevertheless persist. Escape from this state instead requires a rare event, such as immigration or mutation. For example, suppose a mutant $v$ evolves the ability to produce a toxin to which it is immune while the resident species $u$ is not. Both species still compete for the same resources, but the toxin selectively penalizes $u$, increasing the effective competition coefficient $a$ above $1$, whereas the reciprocal coefficient $b$, reflecting only resource competition, remains close to $1$. Such a mutation may therefore drive the system directly into the spinodal regime.

Describing nucleation near the spinodal for non-equilibrium dynamics is therefore relevant not only from a theoretical perspective, as it connects the bistable and monostable regimes, but also for understanding ecological invasions driven by mutations or immigration.

\subsubsection{Critical Nucleus\label{sec:noneqspinnucleus}}

We will show that, like for equilibrium dynamics, for $d<d_\mrm c$ the critical nucleus has a universal shape and involves a small amplitude perturbation around the homogeneous state $\mb u_+^*$, so that $\mb u_\mrm c=\mb u_+^*+\boldsymbol{\delta}\mb u$. We limit ourselves to the generic case $n=3$. Other cases can be treated similarly. 

We consider parameters $\mb p$ of the field close to a set $\mb p_\mrm{sp}$ on the spinodal line. Writing the spinodal limit as $\epsilon=\abs{\mb p-\mb p_\mrm{sp}}\to0$, the critical nucleus is defined by
\begin{equation}
    \mb0=D\Delta\mb u_\mrm c+\mb f_\mb p\qty(\mb u_\mrm c),\quad\mb u_\mrm c\qty(\abs{\mb x}\to+\infty)=\mb u_+^*\text.
    \label{eq:critnuc}
\end{equation}
We thus expand the force field around $\mb u_+^*$,
\begin{equation}
    \mb 0=D\Delta\boldsymbol{\delta}\mb u+\underbrace{\mb f_\mb p\qty(\mb u_+^*)}_{=\mb0}+\evaluate{\partial_j\mb f_\mb p}{\mb u_+^*}\delta u_j+\frac{1}{2}\evaluate{\partial^2_{jk}\mb f_\mb p}{\mb u_+^*}\delta u_j\delta u_k\text,
\end{equation}
where Einstein summation is implied. Henceforth, the derivatives of the force field are understood to be evaluated at $\mb u_+^*$. 

We decompose $\boldsymbol{\delta}\mb u = \sum_i \delta u_i \mb e_i$ in an eigenbasis $\{\mb e_i\}$ of the Jacobian $\mb J_{\mb p_\mrm{sp}}\qty(\mb u_+^*)$, and denote $\lambda_i$ the corresponding eigenvalues. We note $\mb e_1$ the soft eigenvector with $\lambda_1=0$. At second order in $\epsilon$, the critical nucleus verifies
\begin{multline}
    \!\!\!0=D\Delta\delta u_i+\lambda_i\delta u_i+\epsilon M_{ij} \delta u_j+\frac{1}{2}\partial^2_{jk}f_{\mb p_\mrm{sp},i}\delta u_j\delta u_k\text,
\end{multline}
where we introduced
\begin{equation}
\epsilon M_{ij}= \partial_jf_{\mb p,i}-\partial_jf_{\mb p_\mrm{sp},i}\text.
\end{equation}

We expect the soft mode $\mb e_1$ to control the asymptotics, and thus seek a self-consistent scaling in which the perturbation along this direction dominates over the others by writing the amplitudes as
\begin{equation}
\delta u_1=\epsilon U_1, \quad \forall i>1,\ \delta u_i=\epsilon^2U_i\text,
\end{equation}
and rescaling space as $\mb X=\sqrt{\epsilon}\mb x$, so that $\Delta=\epsilon\Delta_{\mb X}$. At leading order, the equation for $U_1$ is
\begin{equation}
    0=D\Delta_{\mb X}U_1+M_{11}U_1+\frac{1}{2}\partial^2_{11}f_{\mb p_\mrm{sp},1}U_1^2\label{eq:u1}
\end{equation}
while for $U_{i>1}$ it reads
\begin{equation}
    0=\lambda_iU_i+M_{i1}U_1\text.
\end{equation}

Since $\mb u_+^*$ is stable for $\epsilon>0$, we have $M_{11}<0$. Assume further that the genericity condition $\partial^2_{11}f_{\mb p_\mrm{sp},1}\neq0$ is verified. Then, Muratov \textit{et al.} proved that bounded solutions to Eq.~(\ref{eq:u1}) exist for $d<d_\mrm c\qty(n)$, with $d_\mrm c=6$ for $n=3$~\cite{muratov_breakup_2004}. This shows that $U_1$ is of order one and $U_i$ is of order at most one for $i>1$, justifying our expansions and our tentative scaling choice.

At leading order in $\epsilon$, the critical nucleus is
\begin{equation}
    \fbox{$\displaystyle\mb u_\mrm c\qty(\mb x)=\mb u_+^*+\epsilon\frac{2\abs{M_{11}}}{\partial^2_{11}f_{\mb p_\mrm{sp},1}}\psi_\mrm c\qty(\sqrt{\epsilon\frac{\abs{M_{11}}}{D}}\mb x)\mb e_1$}\label{eq:spinnuc}
\end{equation}
where $\psi_\mrm c$ is the non-trivial solution to
\begin{equation}
    0=D\Delta\psi_\mrm c-\psi_\mrm c+\psi_\mrm c^2,\quad\psi_\mrm c\qty(\abs{\mb x}\to+\infty)=0\text.
\end{equation}
As an eigenvector, $\mb e_1$ is determined up to a multiplicative constant. Rescaling $\mb e_1$ leads to a corresponding rescaling of $\partial^2_{11}f_{\mb p_\mrm{sp},1}$, such that the final result Eq.~(\ref{eq:spinnuc}) does not depend on that constant.

The eigenvector $\mb e_1$ has a straightforward geometric interpretation in the well-mixed system $\dot{\mb u}=\mb f_\mb p\qty(\mb u)$. At the spinodal, the fixed point $\mb u_+^*$ loses stability by bifurcating with the saddle point $\mb u_\mrm s^*$ belonging to the separatrix between the basins of attraction of $\mb u_+^*$ and $\mb u_-^*$. The fixed points $\mb u_+^*$ and $\mb u_\mrm s^*$ must thus be close before the bifurcation. In fact $\mb e_1$ is the direction from $\mb u_+^*$ to $\mb u_\mrm s^*$ and the easiest way to nucleate is to concentrate the nucleus along that direction.

The spinodal nucleus derived for a vectorial order parameter restores the same kind of universality as in the scalar case: only one eigenvector determines the critical nucleus, and its shape is controlled by the same universal function $\psi_\mrm c$.

\subsubsection{Quasipotential}

We now compute the quasipotential in the spinodal limit directly from its definition as a minimizer of the Freidlin--Wentzell action. The key simplification is that the critical nucleus concentrates on a single eigenvector of $\mb J_{\mb p_\mrm{sp}}$, so that the full vector dynamics reduces, at leading order, to an effective scalar problem. The Freidlin--Wentzell action associated with Eq.~(\ref{eq:rd}) is in general
\begin{equation}
    \frac{1}{4}\int_{\mb u_+^*}^{\mb u_\mrm c}\norm{\mb G^{-\frac{1}{2}}\qty(\mb u)\qty{\partial_t\mb u-\qty[D\Delta\mb u+\mb f\qty(\mb u)]}}^2\mrm dt
\end{equation}
where
\begin{equation}
    \norm{\mb u\qty(\mb x)}^2=\int\abs{\mb u\qty(\mb x)}^2\mrm d\mb x\text.
\end{equation}
At leading order, we can write $\mb u=\mb u_+^*+\delta u_1\mb e_1$, so the action becomes
\begin{multline}
    \frac{1}{4}\abs{\mb G^{-\frac{1}{2}}\qty(\mb u_+^*)\mb e_1}^2\int_{\mb u_+^*}^{\mb u_\mrm c}\\
    \norm{\partial_t\delta u_1-\qty[D\Delta\delta u_1-\epsilon\abs{M_{11}}\delta u_1+\frac{1}{2}\partial^2_{11}f_{\mb p_\mrm{sp},1}\delta u_1^2]}^2\mrm dt\text.
    \label{eq:wentzellfreidlinspin}
\end{multline}

We perform a similar rescaling as above, writing per Eq.~(\ref{eq:spinnuc})
\begin{equation}
    \delta u_1\qty(t,\,\mb x)=\epsilon\frac{2\abs{M_{11}}}{\partial_{11}^2f_{\mb p_\mrm{sp},1}}\psi\qty(T,\,\mb X)\quad\text{with}\quad\mb X=\sqrt{\epsilon\frac{\abs{M_{11}}}{D}}\mb x\text.
\end{equation}
The time scale $T$ is such that all terms inside the norm scale identically, so that $T=\epsilon\abs{M_{11}}t$. Upon this rescaling, the spatial integral in the norm $\norm{\cdot}^2$ contributes a factor $(\mathrm d x/\mathrm d X)^{d}$. Hence, the action reads
\begin{multline}
    \frac{1}{4}\abs{\mb G^{-\frac{1}{2}}\qty(\mb u_+^*)\mb e_1}^2\qty(\epsilon\frac{2\abs{M_{11}}}{\partial_{11}^2f_{\mb p_\mrm{sp},1}})^2\dv{T}{t}\qty(\dv{x}{X})^d\int_0^{\psi_\mrm c}\\
    \norm{\partial_T\psi-\qty[\Delta_{\mb X}\psi-\psi+\psi^2]}^2\mrm dT\text.
\end{multline}
This can be written
\begin{multline}
    \epsilon^{\frac{6-d}{2}}\frac{\abs{M_{11}}^{\frac{6-d}{2}}D^{d/2}\abs{\mb G^{-\frac{1}{2}}\qty(\mb u_+^*)\mb e_1}^2}{\qty(\partial^2_{11}f_{\mb p_\mrm{sp},1})^2}\\
    \times\int_0^{\phi_\mrm c}\norm{\partial_T\psi+\frac{\delta\mathcal E_\mrm{sp}}{\delta\psi\qty(\mb X)}}^2\mrm dT\text,
\end{multline}
where
\begin{equation}
    \mathcal E_\mrm{sp}\qty[\psi\qty(\mb X)]=\int\frac{1}{2}\abs{\boldsymbol{\nabla}\psi}^2+\frac{1}{2}\psi^2-\frac{1}{3}\psi^3\mrm d\mb X\text.
\end{equation}

The problem of minimizing this action is the same as for the equilibrium dynamics
\begin{equation}
    \partial_t\psi=D\Delta\psi-\psi+\psi^2+\eta\qty(t,\,\mb x)\text.
\end{equation}
The MAP for equilibrium dynamics is known and verifies $\partial_T\psi=\frac{\delta\mathcal E_\mrm{sp}}{\delta\psi\qty(\mb X)}$. Therefore, the quasipotential is
\begin{equation}
    \fbox{$\displaystyle U_\mrm c=\epsilon^{\frac{6-d}{2}}\frac{4\abs{M_{11}}^{\frac{6-d}{2}}D^{d/2}\abs{\mb G^{-\frac{1}{2}}\qty(\mb u_+^*)\mb e_1}^2}{\qty(\partial^2_{11}f_{\mb p_\mrm{sp},1})^2}\mathcal E_\mrm{sp}\qty[\psi_\mrm c\qty(\mb X)]$}\text.\label{eq:ucspin}
\end{equation}

Focusing on the trial paths $\mb u=\mb u_+^*+\delta u_1\mb e_1$ is not merely a variational ansatz as an excursion along a stiff direction $i>1$ yields sub-leading contributions to the action. Equation~(\ref{eq:ucspin}) is therefore the quasipotential at leading order in $\epsilon$, not an upper bound on it.

We find a universal scaling $\epsilon^{(d_\mrm c-d)/2}$, with $d_\mrm c\qty(n=3)=6$. It is the same as in equilibrium dynamics, where it is usually established using energetic methods that are unavailable out of equilibrium. The constant $\mathcal E_\mrm{sp}\qty[\psi_\mrm c]$ is universal as well: it depends only on the dimension $d$ and can be computed numerically with the shrinking dimer method, as explained in Sec.~\ref{sec:num}.

The derivation works because, along the MAP, detailed balance is effectively restored after the dynamics collapses onto a single order parameter.

\subsection{Other Cases, Generalizations and Limitations\label{sec:limits}}

We have determined the asymptotic behavior of the quasipotential near the binodal and spinodal in the weak-noise limit. In both regimes, the nucleation probability follows from Eq.~(\ref{eq:defuc}), but it can receive corrections that we now discuss.

Near the spinodal, the main limitation is the weak-noise approximation itself. As in equilibrium spinodal nucleation, the quasipotential vanishes, so fluctuations become increasingly important and critical nuclei fluctuate strongly. These effects are intrinsic to nucleation near the spinodal, but they are difficult to incorporate analytically.

Near the binodal, by contrast, more can be said about the corrections to $U_\mrm c$ and the weak-noise limit.

\subsubsection{Corrections to the Quasipotential}

The quasipotential for spontaneous nucleation, \textit{i.e.} $R_0=0$, scales with the distance $\varepsilon$ to the binodal as
\begin{equation}
    U_\mrm c=\frac{K}{\varepsilon^{d-1}}+o\qty(\frac{1}{\varepsilon^{d-1}})\text,
\end{equation}
obtained because $c$ is of order $\varepsilon$ in the quasipotential Eq.~(\ref{eq:uc}). The constant $K$ is determined by the front properties. It only depends on the position along the binodal, not the distance $\varepsilon$ to it.

This result accounts for the contribution of front motion to the quasipotential. There is also a contribution from the creation of the fronts themselves. We have not computed it analytically because, just as the fronts, it involves the full range of nonlinearities of Eq.~(\ref{eq:rd}), from the metastable state to the stable state. Yet, for small $\varepsilon$, it depends only weakly on $\varepsilon$ because it only involves the shape of fronts rather than their speed. We thus treat that contribution as a constant $\mathcal U$. Hence, close to the binodal, we model the quasipotential starting from $R_0=0$ as
\begin{equation}
    U_\mrm c=\frac{K}{\varepsilon^{d-1}}+\mathcal U\text.\label{eq:ucfit}
\end{equation}
We show in Sec.~\ref{sec:appquasi} that this expression agrees very well with numerical calculations.

\begin{figure}[t!]
    \centering
    \includegraphics[width=0.8\linewidth]{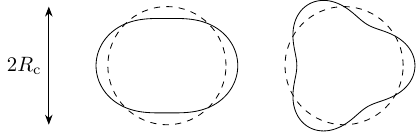}
    \caption{The critical nucleus has a circular shape (dashed). In solid lines, the first (left) and second (right) non-marginal perturbation modes of the critical nucleus shape in two dimensions, whose wavelength scales with $R_\mrm c$.}
    \label{fig5}
\end{figure}

\begin{figure*}[t!]
    \centering
    \includegraphics[width=0.9\textwidth]{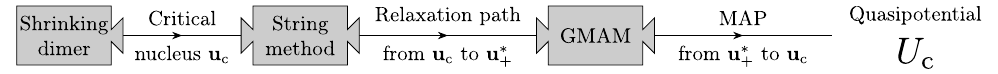}
    \caption{Numerical procedure to calculate the quasipotential $U_c$ entering the nucleation probability Eq.~(\ref{eq:defuc}).}
    \label{fig6}
\end{figure*}

\subsubsection{Finite-Noise Renormalization of Front Motion}

At finite noise, one can reach an effective front dynamics by coarse-graining. On larger length-scales the effective noise reduces and the coarse-grained front equation may itself receive corrections. Fluctuations of the front profile can renormalize the effective speed $c$ in Eq.~(\ref{eq:frontmotion}), and may also modify the statistics of the noise term $\xi$. In equilibrium scalar systems, such renormalizations have been related to asymmetry in the curvature of the potential around either stable state \cite{costantini_asymmetric_2001, kado_microscopic_2024}. This suggests that, beyond the weak-noise limit, the front-based nucleation theory may need renormalized effective parameters rather than the bare ones appearing in Eq.~(\ref{eq:frontmotion}).

Yet, the corresponding renormalization mechanism is not known out of equilibrium, so fluctuations of the front profile may lead to genuinely new effects.

\subsubsection{Prefactor to the Nucleation Probability}

Eq.~(\ref{eq:defuc}) gives only the leading exponential scaling of the nucleation probability in the weak-noise limit. The prefactor can also be computed in the weak-noise limit, in close analogy with the Eyring--Kramers prefactor, although an additional contribution appears because the steady-state distribution is not Boltzmann~\cite{bouchet_generalisation_2016}.

This prefactor involves the hessian of the quasipotential at the critical nucleus, $H_\star$. In equilibrium, where the quasipotential is proportional to the energy, the prefactor is determined by the relaxation rates of the critical nucleus modes. When fluctuations are no longer small, they also modify the barrier itself: the energy $\mathcal E\qty[u_\mrm c]$ in Arrhenius' law is replaced by the free energy~\cite{gardiner_handbook_1983}
\begin{equation}
    \mathcal F\qty[u_\mrm c]=\mathcal E\qty[u_\mrm c]-T\mathcal S\qty[u_\mrm c]\text,
\end{equation}
where the entropic contribution $\mathcal S\qty[u_\mrm c]$ represents the effective phase-space volume accessible from the critical nucleus due to fluctuations once its unstable mode is clamped.

Out of equilibrium, $H_\star$ is not related to the relaxation rates of the eigenmodes in a straightforward way. However, close to the binodal, the eigenmodes of the critical nucleus separate approximately into three classes: perturbations of the internal front structure, perturbations of the nucleus shape, and perturbations of the surrounding metastable state.

Perturbations of the internal front structure, shown in Fig.~\ref{fig9}, relax at rates that are hard to calculate analytically. Yet these rates depend only on the front profile, not on the front speed, and therefore do not depend on $\varepsilon$.

Perturbations of the nucleus shape, on the other hand, do depend on $\varepsilon$ through $R_\mrm c$. As sketched in Fig.~\ref{fig5}, their wavelength is proportional to $R_\mrm c$. They dominate at small $\varepsilon$, because larger critical nuclei relax more slowly: their long-wavelength shape modes are less strongly damped by diffusion.

Crucially, these shape modes relaxes according to Eq.~(\ref{eq:frontmotion}), which is the same whether the underlying mesoscopic model is at equilibrium or not. Therefore, their contribution to $H_\star$ can be adapted from the equilibrium calculation~\cite{langer_statistical_1969}.

\subsubsection{Generalizations\label{sec:generalizations}}

The approach we developed here applied beyond the specific model of Eq.~(\ref{eq:rd}). Close to the binodal, the essential requirement for our coarse-graining procedure to hold is the existence of a stable, gapped front. One may also include different diffusion coefficients for each order-parameter component, encoded by a diffusivity matrix $D_{ij}\Delta u_j$. Other spatial couplings can be added as well, such as terms proportional to $\abs{\boldsymbol{\nabla}\mb u}^2$ as in Active Model A~\cite{caballero_stealth_2020}, or higher-order derivatives such as $\Delta^2\mb u$ to resolve shorter small length scales. We now discuss how these modifications alter Eq.~(\ref{eq:frontmotion}).

With a diffusivity matrix, the coefficient in front of the curvature term changes to an effective diffusivity $\overline D$. No mechanism prevents $\overline D$ from being negative. If both $c$ and $\overline D$ are negative, Eq.~(\ref{eq:frontmotion}) describes the formation of stable nuclei rather than unstable ones, as in equilibrium nucleation. Higher-order spatial derivatives can also generate higher-order curvature terms, leading to richer radius dynamics and possibly to multiple fixed points for the droplet radius.

These features are hallmarks of active phase separation, and they are relevant for systems such as biological condensates or bubbly phase separation~\cite{brangwynne_germline_2009, brangwynne_phase_2013, tjhung_cluster_2018}. Although they have already been identified in conserved systems~\cite{cates_theories_2018, weber_physics_2019, zwicker_intertwined_2022, bauermann_chemical_2022, bauermann_critical_2025}, our framework suggests how they might arise for non-conserved order parameters as well. Comparing the two situations would provide a way to probe the essential ingredients for their appearance. In particular, active-matter models predict distinct notions of surface tension, including negative values~\cite{fausti_capillary_2021}. Studying under which condition $\overline D$ becomes negative could help clarify the origin and generality of these effects.

A dependence of the diffusion coefficient $D$ on the order parameters $\mb u$ can also be considered.
This may be relevant, for instance, in cell tissues, where diffusion models cells pushing each other during proliferation. In this case, $D$ should be taken proportional to the local total growth rate~\cite{kayser_emergence_2018, giometto_physical_2018}.

\section{Numerical Computation of the Quasipotential\label{sec:num}}

The asymptotic theory developed above yields the structure of the nucleation problem out of equilibrium. We now turn to numerical methods, which serve both to test it and to turn it into quantitative predictions for a given model. We have implemented them in two spatial dimensions, to obtain the critical nucleus and to calculate the quasipotential.

Our numerical procedure is outlined in Fig.~\ref{fig6}. We first locate the critical nucleus $\mb u_\mrm c$, then find the relaxation path from $\mb u_\mrm c$ to the metastable state $\mb u_+^*$, which we use as an initial guess to compute the MAP from $\mb u_+^*$ to $\mb u_\mrm c$ and obtain its quasipotential $U_c$. We describe the principle of each method below and provide details in Appendix~\ref{sec:detailnum}.

\subsection{Finding the Critical Nucleus by the Shrinking Dimer Algorithm}

We use the shrinking dimer algorithm (SDA) to find the critical nucleus. This method locates saddle points with one unstable direction in a dynamical system $\dot{\mb u}=\mb b\qty(\mb u)$~\cite{zhang_shrinking_2012}. We denote the saddle point to be found by $\mb u_\mrm c$ and its normalized unstable eigenvector by $\mb V_\mrm c$, defined up to a sign. 

\begin{figure}[t!]
    \centering
    \includegraphics[width=\linewidth]{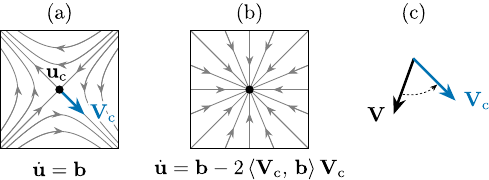}
    \caption{Principle of the shrinking dimer algorithm. (a) Original flow field $\mb b$ close to the saddle point $\mb u_\mrm c$ with one unstable direction $\mb V_\mrm c$. (b) Modified flow after reflection along $\mb V_\mrm c$. (c) The guess $\mb V$ converges to $\mb V_\mrm c$.}
    \label{fig7}
\end{figure}

The SDA achieves convergence of its estimate $\mb u$ to $\mb u_\mrm c$ by considering a different flow, for which $\mb u_\mrm c$ becomes a stable attractive fixed point,
\begin{equation}
    \dot{\mb u}=\mb b\qty(\mb u)-2\left\langle\mb V_\mrm c,\,\mb b\qty(\mb u)\right\rangle\mb V_\mrm c\text.\label{eq:sdaideal}
\end{equation}
In Eq.~(\ref{eq:sdaideal}), the direction of motion along $\mb V_\mrm c$ is reflected to ensure stability. In Fig.~\ref{fig7}, we show the phase portraits of the original flow and the modified one Eq.~(\ref{eq:sdaideal}) close to $\mb u_\mrm c$.

However, $\mb V_\mrm c$ is not known in advance, so the SDA instead uses
\begin{equation}
    \dot{\mb u}=\mb b\qty(\mb u)-2\left\langle\mb V,\,\mb b\qty(\mb u)\right\rangle\mb V\text,
\end{equation}
where $\mb V$ is an estimate of $\mb V_\mrm c$ evolved jointly with $\mb u$ according to
\begin{equation}
    \dot{\mb V}=\mb J\qty(\mb u)\mb V-\left\langle\mb V,\,\mb J\qty(\mb u)\mb V\right\rangle\mb V
    \text,
\end{equation}
with $\mb J\qty(\mb u)$ the Jacobian of $\mb b$ at $\mb u$. The first term aligns $\mb V$ with $\mb V_\mrm c$, since $\mb J\qty(\mb u_\mrm c)$ shrinks every direction except $\mb V_\mrm c$, which it expands. The second term enforces the normalization of $\mb V$ by projecting the dynamics orthogonally to $\mb V$.

\subsection{Initial Condition for the GMAM by the String Method}

We then determine the relaxation path from $\mb u_\mrm c$ to the metastable state $\mb u_+^*$. At equilibrium, this path coincides with the MAP. This is not true out of equilibrium, but it still provides a good initial guess for convergence toward the MAP.

The relaxation path is the deterministic trajectory connecting $\mb u_\mrm c$ to $\mb u_+^*$. It is thus a solution of the field equation~(\ref{eq:rd}), but it cannot be determined by direct simulation of Eq.~(\ref{eq:rd}). Indeed, both endpoints are fixed points of the dynamics, so a direct simulation initialized at either state would remain there indefinitely. Mathematically, this path is a homoclinic orbit, and the time needed to cross it under Eq.~(\ref{eq:rd}) is infinite.

We therefore use the string method~\cite{e_string_2002, e_simplified_2007} to find the relaxation path. It does so iteratively, starting from a discrete sequence of states $\mb u_i^0$, called images, which we choose to interpolate linearly between $\mb u_+^*$ and $\mb u_\mrm c$. At iteration $t$, we write the images $\mb u_i^t$. They evolve by two alternating steps:
\begin{enumerate}
    \item \textit{Evolution step.} Each image is evolved according to the deterministic flow $\dot{\mb u}_i=\mb b\qty(\mb u_i)$ for a fixed time interval, mapping $\mb u_i^t$ to $\Tilde{\mb u}_i^t$;
    \item \textit{Interpolation step.} New images $\mb u_i^{t+1}$ are generated by interpolating the $\Tilde{\mb u}_i^t$ so that the resulting $\mb u_i^{t+1}$ are evenly spaced between $\mb u_+^*$ and $\mb u_\mrm c$. This ensures that the path is accurately resolved, especially in regions where the flow $\mb b$ is strong.
\end{enumerate}

Applying the string method between the two homogeneous states $\mb u_+^*$ and $\mb u_-^*$ would yield a path passing through the critical nucleus $\mb u_\mrm c$ without requiring us to find it first using the SDA. However, many images would be used to sample the path between $\mb u_\mrm c$ and the stable state $\mb u_-^*$, which is not useful because this part of the path contributes no action.

\begin{figure*}[t!]
    \centering
    \includegraphics[width=.95\textwidth]{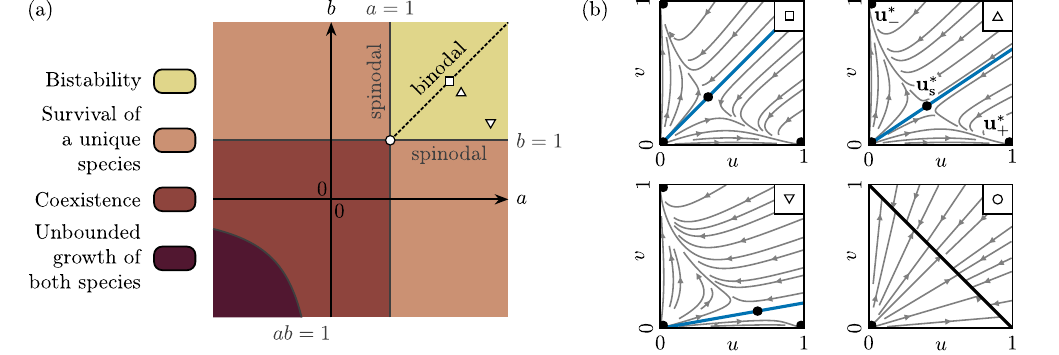}
    \caption{(a) Long-time behavior of the well-mixed Lotka--Volterra model as a function of inter-species interactions $a$ and $b$. We focus on the bistability region $a, b>1$. We call spinodal the lines at $a=1$ and $b=1$ at which one of the species becomes linearly unstable. We call binodal the line along $a=b$ (dashed), on which no species has a net competitive advantage. For selected values of the interaction parameters (white symbols), we show in (b) their phase portraits: $\square ~a=b=2$, $\triangle~a=2.2, b=1.8$, $\bigtriangledown~a=2.7,b=1.3$, $\circ~a=b=1$. The fixed points are located by black dots or a black line. The separatrix between two basins of attraction is shown in blue.}
    \label{fig8}
\end{figure*}

\subsection{GMAM Algorithm\label{sec:numgmam}}

The final step of our numerical procedure is to obtain the MAP from the relaxation path. We use the Geometric Minimum Action Method (GMAM)~\cite{heymann_pathways_2008}. The GMAM proceeds similarly to the string method: it represents the path by discrete images and alternates dynamical and interpolation steps, but the evolution follows a different flow.

This flow is proportional to the gradient descent flow of the geometric action $\mathcal A_\mrm g$. This action differs from the Freidlin--Wentzell action $\mathcal A$, but it has the same minima and minimizers. The geometric action is
\begin{multline}
    2\mathcal A_\mrm g\qty[\mb u\qty(s)]=\\
    \int\norm{\mb G^{-\frac{1}{2}}\frac{\partial\mb u}{\partial s}}\norm{\mb G^{-\frac{1}{2}}\mb b\qty(\mb u\qty(s))}-\left\langle\mb G^{-1}\frac{\partial\mb u}{\partial s},\,\mb b\qty(\mb u\qty(s))\right\rangle\mrm ds\text,
\end{multline}
which, crucially, does not depend on the parametrization $s$ of the path $\mb u\qty(s)$, unlike $\mathcal A$. This makes its gradient descent flow compatible with the interpolation step, which is necessary to sample the path accurately over its full span. Note that the evolution step in the string method acts independently on each image, whereas the gradient of $\mathcal A_\mrm g$ couples the images in GMAM.

Other methods exist to calculate the quasipotential~\cite{e_minimum_2004, zakine_minimum-action_2023}. We have chosen GMAM for its simplicity and ability to scale to high-dimensional systems. Scalability is particularly important close to the binodal. Indeed, in this limit, the critical radius diverges while the front width remains constant, so both scales must be resolved accurately to obtain the quasipotential. Close to the spinodal, the critical nucleus also diverges, but it makes up the only length scale present in the MAP, so the spatial discretization can be scaled with the critical nucleus.

We implemented GMAM in a semi-implicit, pseudo-spectral framework optimized for high-dimensional computations. All computations involving the images independently were carried out in parallel (see Appendix \ref{sec:detailnum}). This allowed us to perform simulations with $400$ images on a grid of lateral size $512$, corresponding to gradient descent in a space of dimension $2\times400\times512^2=2\times10^8$.

\section{Fronts In The Lotka--Volterra System\label{sec:frontslv}}

As a first test of our theory, we study the deterministic properties of fronts in the LV model, namely their stability, their shape and their speed. Firstly, we recall results about the phase portrait of the LV system that will be useful to understand the properties of its fronts.

\subsection{Phase Portrait of the Lotka--Volterra Model\label{sec:portraitlv}}

In general, the LV model has four fixed points,
\begin{equation}
    \mb0,\,\mb u_+^*=\mqty(1\\0),\,\mb u_-^*=\mqty(0\\1),\,\mb u_\mrm s^*=\frac{1}{ab-1}\mqty(a-1\\b-1)\text.
\end{equation}
The first is always unstable and is of no interest to us. The stability of the others depends on $a$ and $b$ and determines the long-time behavior of the system, see Fig.~\ref{fig8}(a).

We focus on the bistable regime, $a,b>1$ such that both $\mb u_+^*$ and $\mb u_-^*$ are linearly stable, and $\mb u_\mrm s^*$ is a saddle point that separates their basins of attraction. There, the long-time behavior depends on the initial conditions~\cite{strogatz_nonlinear_2019}. If $\qty(b-1)u\qty(0)>\qty(a-1)v\qty(0)$ the system converges to $\mb u_+^*$, otherwise to $\mb u_-^*$. We call binodal the line $a=b$ on which the volume of both basins of attraction is equal, see the top left portrait in Fig.~\ref{fig8}(b).

The lines $a=1$ and $b=1$ mark the boundaries of the bistable region, and are thus spinodal limits. As $b\to1$, $\mb u_\mrm s^*$ approaches $\mb u_+^*$ and the basin of attraction of $\mb u_+^*$ shrinks, see Fig.~\ref{fig8}(b). Thus, when $b\gtrsim1$, $\mb u_+^*$ is still stable to infinitesimal perturbations, but even small perturbations can push the system into the basin of attraction of $\mb u_-^*$. When $b<1$, $\mb u_+^*$ is linearly unstable: introducing any amount of the species $v$ leads to its invasion and the disappearance of $u$. The situation is symmetric as $a\to1$.

When $a$ and $b$ approach $1$ jointly, however, the situation is different. This limit corresponds to ecological neutrality, where interspecific and intraspecific interactions are identical. In this case, $u$ and $v$ can be viewed as an arbitrary partition of a single species. The system is stationary whenever the carrying capacity is shared as $u+v=1$, while the relative frequency $f=u/\qty(u+v)$ can take any value. Thus, the states $\qty(u=f,\,v=1-f)$ with $0\leq f\leq1$ form a continuous line of fixed points, as illustrated in the bottom right portrait of Fig.~\ref{fig8}(b).

\subsection{Stability of the Fronts}

We now turn to fronts in the LV system. First, we show numerically that fronts are linearly stable and gapped, which are conditions for our nucleation theory to hold. To find fronts, we performed direct simulations of Eq.~(\ref{eq:rd}) in $d=1$ without noise, with a step initial condition $\mb u\qty(t=0,\,x<0)=\mb u_-^*$ and $\mb u\qty(t=0,\,x>0)=\mb u_+^*$. The front solution $\mb u_\mrm{fr}\qty(x)$ is extracted from the stationary profile of $\mb u$. We provide numerical details in Appendix~\ref{sec:numfronts}. Our procedure relies on fronts being stable attractors, which we confirm. To check their spectral properties, we define the operator $\mathcal L$ as the linearization of Eq.~(\ref{eq:rd}) around the front, namely
\begin{equation}
    \mathcal L\qty[\boldsymbol{\delta}\mb u]=D\frac{\partial^2\boldsymbol{\delta}\mb u}{\partial x^2}+\mb J_\mb p\qty(\mb u_\mrm{fr})\cdot\boldsymbol{\delta\mb u}\text,
\end{equation}
with $\mb J_\mb p$ the Jacobian of $\mb f_\mb p$. The perturbation modes are eigenvectors of $\mathcal L$, with eigenvalues $\Lambda_k$ ordered in ascending order from $k=1$. They are proportional to the relaxation rate of perturbations along the corresponding mode.

\begin{figure}[t!]
    \centering
    \includegraphics[width=0.9\linewidth]{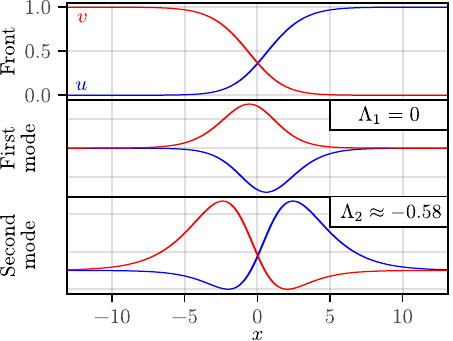}
    \caption{Front in the LV system on the binodal ($a=b=2$) and its first two perturbation modes. The front was found numerically and the modes were found by numerically diagonalizing the operator $\mathcal L$. No vertical scale is shown for eigenvectors because they are defined up to a multiplicative constant.}
    \label{fig9}
\end{figure}

In Fig.~\ref{fig9}, we show a front for parameters on the binodal, together with its first two perturbation eigenmodes and corresponding eigenvalues $\Lambda_1$ and $\Lambda_2$. Both eigenvalues are non-positive, confirming that the front is stable. The first eigenmode is the translational mode $-\mathrm d \mb u_\mrm{fr} /\mathrm dx$ and is marginal, $\Lambda_1=0$, as a consequence of the translational invariance of Eq.~(\ref{eq:front}). The second eigenmode governs the slowest decaying deformation of the front. Its shape shows that it primarily corresponds to a change in the front width: perturbations along this mode compress or broaden the transition region between the two homogeneous states. Hence, the front width is the slowest relaxing internal degree of freedom. We confirm the existence of a gap $\Lambda_2-\Lambda_1$ between the first two modes. We checked that this gap is not a numerical artifact: the eigenvalues are independent of both the simulation-domain length and the spatial discretization of the front (see Appendix~\ref{sec:numfronts}).

Fronts are thus stable and gapped. We turn to their dependence on the interaction parameters $a$ and $b$.

\begin{figure*}[t!]
    \centering
    \includegraphics[width=\textwidth]{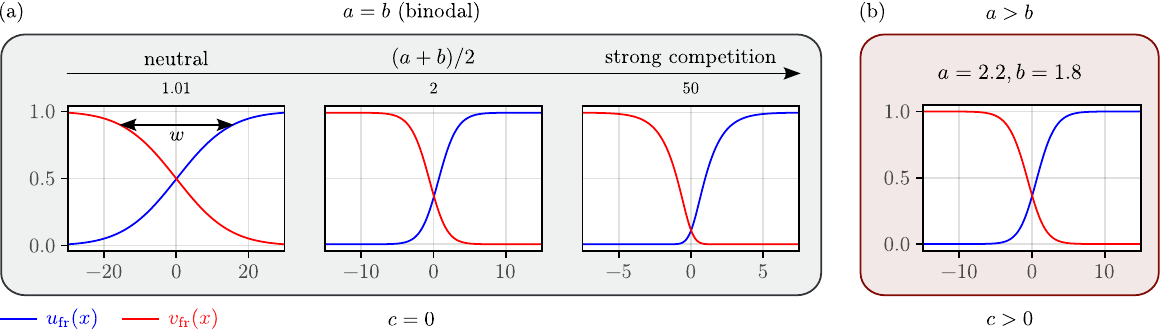}
    \caption{Shape of the fronts joining the two stable states $\mb u_+^*=\qty(u=1,\,v=0)$ and $\mb u_-^*=\qty(u=0,\,v=1)$ in the LV model. (a) Along the binodal line $a=b$, species are equally good competitors, thus fronts have zero speed $c=0$. From left to right, the interaction parameters increase from $(a+b)/2 = 1.01$ to $2$ and $50$, \textit{i.e.} from the neutral to strongly competing limit. In panel (b), the parameters are $a=2.2,b=1.8$ such that $(a+b)/2 = 2$, as the center plot of (a), but now $a>b$. The shape of the front is unchanged compared to the binodal case, but the front now moves at a non-zero speed $c>0$. Units of space are such that $D=1$, and $x=0$ is arbitrarily chosen to be the center of the front.}
    \label{fig10}
\end{figure*}

\subsection{Shape of the Fronts: Profile, Width and Internal Structure}

By symmetry, it is clear that $c=0$ when $a=b$, which characterizes the binodal line. Since increasing $a$ hinders the growth of $u$, we get $c>0$ whenever $a>b$. In the following, we always take $a\geq b$, so that $v$ is a better competitor than $u$, meaning that $\mb u_-^*$ is able to invade a region occupied by $\mb u_+^*$ with speed $c\geq0$. In particular, this means that large inocula of $v$ in a background of $u$ will grow, but any finite inoculum of $u$ in a region dominated by $v$ will decay.

\textit{Front profile.\,---} We obtained numerically the front profiles for various values of $a$ and $b$. We also compute their shape analytically in the limits of neutrality $a,\,b\to1$ and strong competition $a=b\to+\infty$, as detailed in Appendix \ref{sec:fronts}. Figure \ref{fig10} shows some representative fronts, both on the binodal $a=b$ and off. Together with the analytical calculations, we find that front profiles are mostly determined by $\frac{1}{2}\qty(a+b)$ and depend only weakly on $a-b$. When $a$ and $b$ are close, the quantity $\frac{1}{2}\qty(a+b)$ indicates the position along the binodal line. It determines the shape and internal front structure. Instead, $a-b$ captures the distance to the binodal, and determines the front speed.

\begin{figure}[b!]
    \centering
    \includegraphics[width=\linewidth]{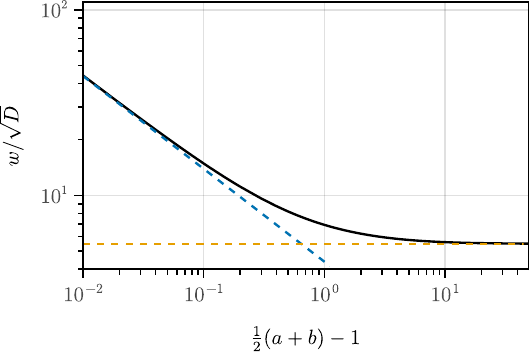}
    \caption{Front width $w$ along the binodal line ($a=b$) as a function of the interaction strength, from the neutral (left) to strongly competing limit (right). The data is measured via direct numerical simulations of the fronts (black line). In dashed, the asymptotes obtained analytically in the neutral (blue, Eq.~(\ref{eq:frontneutralxi})) and strongly competing limits (yellow, Eq.~(\ref{eq:frontstrongw})).}
    \label{fig11}
\end{figure}

\textit{Front width.\,---} We define the front width as the distance $w$ between the positions where the front reaches $v_\mrm{fr}=0.9$ and $u_\mrm{fr}=0.9$. We see from Fig.~\ref{fig10} that the front width $w$ changes along the binodal. In Fig.~\ref{fig11}, we report numerical results for the front width as a function of the competition strength $\frac{1}{2}\qty(a+b)-1$, where the shift allows to study both neutral (to the left) and strong competition (to the right) limits. From strong competition to the neutral limit, the front width increases. It even diverges upon approaching ecological neutrality. In this limit, the well-mixed system is stationary for $u+v=1$, irrespective of $f$. The consequence on the spatial system is that once the total abundance $u+v$ has relaxed to $1$ at every point, diffusion remains the only process shaping the dynamics, which spreads fronts outwards indefinitely. The situation is similar to that of physical systems with a continuous symmetry, such as the XY model, which do not exhibit fronts. We analytically compute the asymptotic behavior of the front width in both limits. In the neutral limit, we find that it diverges as $w/\sqrt{D}=\mathcal O\qty(1/\sqrt{a+b-2})$. In the strong-competition limit, the width tends to a constant. See Appendix~\ref{sec:fronts} for detailed calculations of both asymptotes. The asymptotic predictions and numerical data show great agreement.

\begin{figure}[t!]
    \centering
    \includegraphics[width=\linewidth]{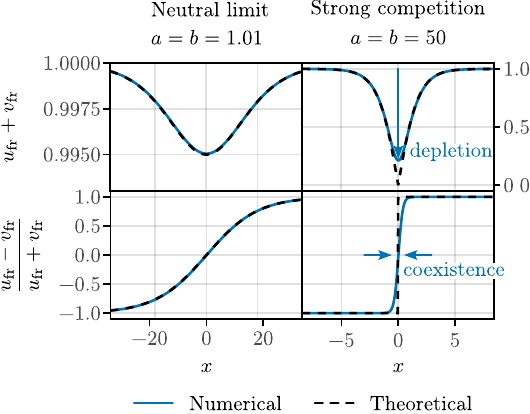}
    \caption{(Top) Total abundance $u_\mrm{fr}+v_\mrm{fr}$ along the front close to the neutral (left) and strong competition (right) limits. For strongly competing species, a depletion region emerges at the center of the front, where both species are suppressed $u_\mrm{fr}+v_\mrm{fr} \ll 1$ and even extinct in the limit $a=b\rightarrow\infty$ (dashed line). Such depletion is barely noticeable close to neutrality. (Bottom) The relative abundance difference goes from $-1$ to $1$ over the region where both species coexist. This coexistence region is very broad close to neutrality, decreases for increasing competition and eventually vanishes in the limit $a=b\rightarrow\infty$ (step dashed line). }
    \label{fig12}
\end{figure}

\textit{Front internal structure.\,---}Moreover, the front has an internal structure on a length scale smaller than $w$. Figure~\ref{fig12} shows the same fronts as Fig.~\ref{fig10}, in both the neutral and strong-competition limits, but using different variables. Instead of plotting $u_\mathrm{fr}$ and $v_\mathrm{fr}$ separately, we show the total abundance $u_\mathrm{fr}+v_\mathrm{fr}$ and the relative abundance difference $(u_\mathrm{fr}-v_\mathrm{fr})/(u_\mathrm{fr}+v_\mathrm{fr})$. In the neutral limit (left panel), both quantities evolve on the same scale $w$. As mentioned above, the total abundance is close to $1$ everywhere. By contrast, in the limit of strong competition (right panel), the total abundance evolves on the scale of $w$ but the relative abundance difference evolves on a smaller scale, which vanishes in the large competition limit, $\frac{1}{2}\qty(a+b)\to\infty$. 
The theoretical curves in the left panel of Fig.~\ref{fig12} are computed as an expansion around the neutral point $a=b=1$. In the right panel, they are computed for the limit $a=b\rightarrow\infty$. Both are derived in Appendix~\ref{sec:fronts}.

On the large scale, the strong dip (top panels) reveals that the front is a region of depletion for the total abundance of individuals. Indeed, when competition is large, it is harder for either species to coexist with the other long enough to diffuse on large scales. Only over the region where the relative abundance difference is meaningfully non-zero do the species coexist (bottom panels), with abundances going to $0$ as $\frac{1}{2}\qty(a+b)\to\infty$. \\

Giometto \textit{et al.} considered more realistic models of spatially competing bacterial strains and found numerically qualitatively similar front shapes, and a depletion region is clearly visible in their experiments~\cite{giometto2021}. This suggests that our results can be representative of real situations. Yet, competition is often modeled using an effective scalar order parameter, such as the relative frequency $f$, which cannot capture the depletion region~\cite{lavrentovich_asymmetric_2014, tanaka_spatial_2017, lavrentovich_nucleation_2019}. A vectorial description is therefore essential to describe this feature. Moreover, the depletion region is not merely a detail of the front profile: we will show that it strongly influences front properties at large competition strength. More complex local terms $\mb f_\mb p$ and a greater number of variables, such as more species or toxin concentrations, needed to model quantitatively the above experiments can easily be included in our framework.

\subsection{Speed of the Fronts\label{sec:susceptibility}}

We now discuss the predictions of Eq.~(\ref{eq:speed}) for the speed $c$ of a flat front in the LV system close to the binodal. We map given parameters $\mb p=\mqty(a&b)^\mathsf T$ to $\mb p_\mrm b=\mqty(\frac{1}{2}\qty(a+b)&\frac{1}{2}\qty(a+b))^\mathsf T$, so that the distance to the binodal is $\varepsilon\sim a-b$. Then, using Eq.~(\ref{eq:deltafgeneral}), the driving force for flat fronts is
\begin{equation}
    \Delta\mb f\qty(\mb u_\mrm{fr}^\mrm b)=\frac{a-b}{2}u_\mrm{fr}^\mrm bv_\mrm{fr}^\mrm b\mqty(-1\\1)\text.\label{eq:deltaf}
\end{equation}
Fronts are thus driven by the competitive advantage of the better competitor in regions where both species are present. As a consequence, their speed is proportional to $a-b$ for $a\approx b$, prompting us to define the susceptibility 
\begin{equation}
    \chi=\lim_{a-b\to0}\frac{c}{a-b}\text.
\end{equation}

\begin{figure}[t!]
    \centering
    \includegraphics[width=\linewidth]{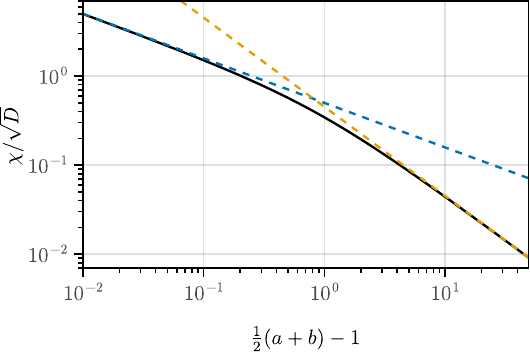}
    \caption{Susceptibility $\chi$ of fronts as a function of the competition strength. The plotted quantity is the dimensionless ratio $\chi/\sqrt{D}=c/\qty(a-b)\sqrt{D}$ in the binodal limit $a-b\to0$. The black line is obtained by numerically evaluating Eq.~(\ref{eq:speed}), while the dashed lines are the analytical asymptotic limits given by Eqs.~(\ref{eq:frontneutralxi}) and (\ref{eq:speedstrong}).}
    \label{fig13}
\end{figure}

The value of $\chi$ for varying values of $\frac{1}{2}\qty(a+b)$ is plotted in Fig.~\ref{fig13}. We find that fronts in highly competitive systems have smaller susceptibility than those in less competitive systems. Given the profile of fronts, this behavior is readable from the derived expression for the flat front speed $c$, Eq.~(\ref{eq:speed}). Indeed, fronts for large $\frac{1}{2}\qty(a+b)$ feature a characteristic depletion of the total abundance and a very small coexistence region, as illustrated in Fig.~\ref{fig12}. The region over which both species have non-zero abundances is thus very narrow, and the abundances are very low. Since Eq.~(\ref{eq:deltaf}) predicts that competition occurs in this region, there is less opportunity for either species to effectively manifest its superiority. As a consequence, the susceptibility is lower.

The situation is entirely converse in the limit of neutrality $a,\,b\to1$. The region where both species coexist is then very large, and the total abundance does not decrease, therefore competition can be most efficient.

Analytical results detailed in appendix \ref{sec:fronts} complete this analysis and yield that for $a,\,b\to1$, the speed varies like $c\propto\qty(a-b)/\sqrt{a+b-2}$, whereas for large $\frac{1}{2}\qty(a+b)$, it goes as $c\propto\qty(a-b)/\qty(a+b)$. Although they feature higher susceptibility, smaller values of $\frac{1}{2}\qty(a+b)$ allow for a smaller range of $a-b$, limiting the maximum speed of the front. Ultimately, the value of front speed $c$ is a trade-off between the forcing strength $a-b$ and its susceptibility as dictated by the value of $\frac{1}{2}\qty(a+b)$.

In the LV model, we have assumed that interactions are purely local. The results of our analysis could change if they were mediated in space, for instance by diffusion of a resource or a toxin. However, no matter the underlying model, we have shown that our method provides interpretable results for the speed of fronts, which can be related to the profile of abundances across the front. From a physical perspective, this interpretability is particularly precious since the front speed takes on the conceptual role of the volumetric bulk energy difference.

\section{Nucleation In The Lotka--Volterra Model\label{sec:applicationlv}}

We now explore the consequence of these front characteristics in the Lotka--Volterra model on the nucleation properties. We first detail our results about the critical nuclei and then show how they can be used to understand nucleation in confined systems. Finally, we consider the quasipotential under different noise terms.

\subsection{Critical Nuclei and Effects of Confinement}

\subsubsection{Critical Nuclei in Infinite Space}

\begin{figure}
    \centering
    \includegraphics[width=\linewidth]{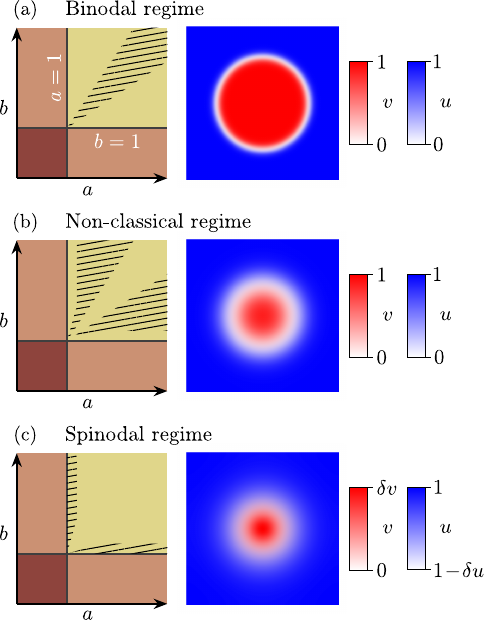}
    \caption{Critical nuclei in the Lotka--Volterra model found using the shrinking dimer algorithm and regimes of interaction parameters (hatched $a,\,b$) that feature similar nuclei. (a) Close to the binodal $a=b$, the critical nucleus resembles a droplet with interfaces that are thin compared to its radius. This is the non-equilibrium analogue of the Classical Nucleation Theory (CNT) regime. (b) Away from the binodal, the interface width is no longer small compared to the radius, and the order parameters are no longer homogeneous inside the droplet, the analogue of non-classical nucleation. (c) Close to the spinodals $a=1$ and $b=1$, the critical nucleus becomes increasingly wide and its amplitude decreases, as in equilibrium spinodal nucleation.}
    \label{fig14}
\end{figure}

We use the numerical methods presented in Sec.~\ref{sec:num} to obtain the critical nucleus as a function of interaction parameters $a$ and $b$ in the Lotka--Volterra model. We show  the resulting critical nuclei in Fig.~\ref{fig14}, for parameters close to the binodal, away from it, and close to the spinodal limit. In all cases, the domain size $L$ is chosen much larger than the critical nucleus, so that the nucleus is unaffected by the boundaries of the numerical domain and can be considered representative of the critical nucleus in an infinite system.

In panel (a), the critical nucleus is shown close to the binodal $a=b$ and resembles a droplet with sharp edges, consistent with our prediction that it can be described using curved fronts. The hatched region in the figure is an approximate representation of the domain where we find $R_\mrm c\gtrsim w$ numerically and where our non-equilibrium nucleation theory should hold.

In the Supplementary Material (Videos 1 and 2), we illustrate the dynamical role of the critical nucleus in the LV model. Initial droplets larger than the critical nucleus grow, whereas smaller ones shrink. Moreover, the same critical nucleus is obtained even when starting from non-circular inocula: the configuration separating growth from decay first relaxes toward the critical nucleus before remaining stationary.

In panel (c), we show a critical nucleus in the spinodal limit $b\to1$. According to Eq.~(\ref{eq:spinnuc}), the critical nucleus is
\begin{equation}
    \left\{\begin{array}{l}
        u_\mrm c\qty(\mb x)=1-a\frac{b-1}{a-1}\psi_\mrm c\qty(\sqrt{\frac{b-1}{D}}\mb x)\\
        v_\mrm c\qty(\mb x)=\frac{b-1}{a-1}\psi_\mrm c\qty(\sqrt{\frac{b-1}{D}}\mb x)
    \end{array}\right.\text,
\end{equation}
thereby defining $\delta u=a\frac{b-1}{a-1}\psi_\mrm c\qty(\mb0)$ and $\delta v=\frac{b-1}{a-1}\psi_\mrm c\qty(\mb0)$, which determine the range of colors in the heatmap. The numerical determination of the critical nucleus is consistent with our theoretical predictions. Checking the assumptions made along the derivation of Eq.~(\ref{eq:spinnuc}) in the case of the LV model requires that $b-1\ll\min\qty(1,\,1-1/a)$. The corresponding domain is hatched in Fig.~\ref{fig14}(c).

Panel (b) shows a region akin to non-classical nucleation, between the domains of validity of CNT and spinodal nucleation, where the critical nucleus is an inhomogeneous droplet whose edges are thick compared to its radius, for which no analytic prediction is available. Its exact shape is expected to depend on the details of $\mb f_\mb p$ and interpolates between the nuclei in the binodal and spinodal regime.

Therefore, we confirm that our theoretical predictions for critical nuclei in different regimes are correct. We now use them to discuss the effects of confinement, which affect many experimental realizations. Physical boundaries can limit the space available for a nucleus to grow, while in discrete systems such as arrays of wells, the finite number of sites sets the system size directly. In both cases, the system size can become comparable to the size of the critical nucleus and thereby affect the nucleation process.

\subsubsection{Effects Of confinement\label{sec:confinement}}

We have shown that the critical nucleus grows very large close to the binodal $a=b$. So far, we have increased the numerical domain size to ensure that it remains larger than the critical nucleus (as in Fig.~\ref{fig14}). However, experiments are often confined to a finite, fixed domain~\cite{copeland_spatial_2024}. What happens when the parameters are such that the critical diameter $2R_\mrm c$ exceeds the domain size? To investigate the effects of confinement on nucleation, we study how the critical nucleus evolves as the linear domain size $L$ is decreased. We model boundaries by imposing Neumann boundary conditions in the shrinking dimer procedure. Ecologically, this corresponds to reflecting boundaries, where individuals reaching the domain boundary are reflected back into it.

\begin{figure}[t!]
    \centering
    \includegraphics[width=\linewidth]{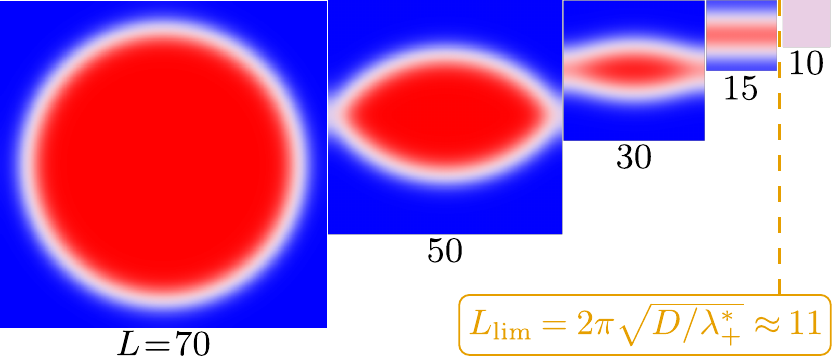}
    \caption{Effect of confinement on the shape of the critical nucleus for $a=2.05$, $b=1.95$. In the unconfined limit of large $L$, the critical radius is $R_{\mrm c}^\infty = 31$. From left to right: critical nucleus obtained for domains of linear size $L = 70, 50, 30, 15, 10$, with Neumann boundary conditions. We show the existence a limit system size, $L_\mrm{lim}$, below which the critical nucleus disappears. The interaction parameters and diffusion coefficient are $a=2.05$, $b=1.95$ and $D=1$.}
    \label{fig15}
\end{figure}

In Fig.~\ref{fig15}, we show the critical nucleus for several domain sizes at fixed parameters $a=2.05$, $b=1.95$, in the binodal nucleation regime. In the unconfined limit of large $L$, the critical radius is $R_{\mrm c}^\infty = 31$. For $L=70$, which is slightly larger than $2R_{\mrm c}^\infty$, the critical nucleus remains circular, as in the large-$L$ limit. However, its shape changes strongly as $L$ is decreased below $2R_{\mrm c}^\infty$.

More precisely, the critical nucleus evolves from a circular shape to an elongated shape delimited by two symmetric circular arcs with radius of curvature $R_{\mrm c}^\infty$. Indeed, the dynamics of fronts in the bulk is still described by Eq.~(\ref{eq:frontmotion}), so that the same curvature is required for the front to remain stationary. However, a full disk of radius $R_{\mrm c}^\infty$ no longer fits inside the domain, and the nucleus therefore consists of two arcs connected at the boundaries. Similar results have been found using CNT for models at equilibrium but studies have mostly focused on systems with a conserved order parameter, in which the elongated droplet shape appeared not as a saddle point for nucleation but for the transition between different shapes of a persistent droplet~\cite{leung_geometrically_1990, neuhaus_2d_2003, schrader_simulation_2009, macdowell_computer_2011}.

As the domain size decreases further, the critical nucleus eventually becomes an elongated strip. At this stage, the nucleus depends on only one coordinate and resembles two adjacent fronts, each extending over a length $L/2$. For the values of $L$ where the critical nucleus is anisotropic, it is doubly degenerate and can be aligned with either principal axis of the domain.

Moreover, we predict the existence of a limiting domain size $L_\mrm{lim}$ below which nucleation occurs homogeneously. While it is expected that the critical nucleus approaches the well-mixed saddle point $\mb u_\mrm s^*$ in the small-$L$ limit, the existence of a sharp threshold for this transition is not obvious. To see this, we assume that the critical nucleus is close to $\mb u_\mrm s^*$ for small $L$ and expand $\mb u_\mrm c=\mb u_\mrm s^*+\boldsymbol{\delta}\mb u\qty(\mb x)$. Then, Eq.~(\ref{eq:critnuc}) becomes
\begin{equation}
    \mb0=D\Delta\boldsymbol{\delta}\mb u+\mb J_\mb p\qty(\mb u_\mrm s^*)\cdot\boldsymbol{\delta}\mb u\text.
\end{equation}
Each eigenvector of the Jacobian $\mb J_\mb p\qty(\mb u_\mrm s^*)$ yields a solution with a wavenumber proportional to the corresponding eigenvalue. For sufficiently small $L$, none of these oscillating solutions is compatible with the boundary conditions, so that $\boldsymbol{\delta}\mb u=\mb0$. The first mode to develop has the smallest wavelength and is thus associated with the largest eigenvalue of $\mb J_\mb p\qty(\mb u_\mrm s^*)$, which we denote by $\lambda_+^*$. It develops when its wavelength matches $L$, yielding
\begin{equation}
L_\mrm{lim}=2\pi\sqrt{D/\lambda_+^*}\text.
\end{equation}
For the parameters considered here, we find $L_\mrm{lim}\approx11$, consistent with the numerical results shown in Fig.~\ref{fig15}.

\subsection{Quasipotential with Langevin Noise\label{sec:appquasi}}

We now test our predictions for the quasipotential close to the binodal and close to the spinodal against direct numerical calculations of $U_c$, and then draw ecological conclusions from them.

We start with Langevin noise, $\mb G\qty(\mb u)=\mb 1$, for which the noise term of Eq.~(\ref{eq:rd}) is additive. This choice is not ecologically realistic, as it can drive abundances negative, but it provides the simplest test of our methods and a reference for the more realistic noise considered in Sec.~\ref{sec:quasi}, which we find to display the same trends.

\subsubsection{Agreement Between Theory and Numerical Calculations}

\begin{figure}
    \centering
    \includegraphics[width=\linewidth]{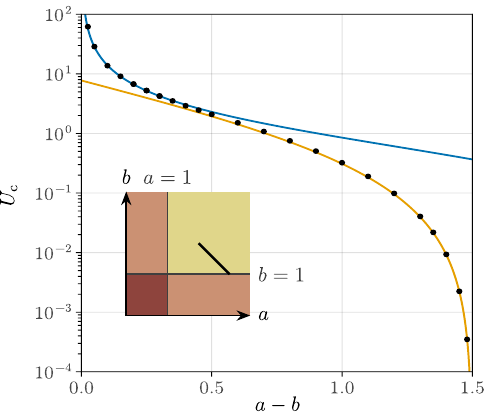}
    \caption{Quasipotential $U_\mrm c$ for varying $a-b$ across the domain of bistability, from the binodal $a=b$ to the spinodal line $b=1$, along the thick black line in the inset. The noise is Langevin, $\frac{1}{2}\qty(a+b)=1.75$, $D=1$.. Black dots are results from the GMAM algorithm. The blue line results from Eq.~(\ref{eq:ucfit}) with $K$ calculated numerically independently from the GMAM data and $\mathcal U$ fitted to the GMAM data close to the binodal. The orange line is the result of Eq.~(\ref{eq:ucspinlv}). }
    \label{fig16}
\end{figure}

Starting with a fixed value of $\frac{1}{2}\qty(a+b)$, we vary $a-b$ and numerically compute the value of $U_\mrm c$ using the GMAM algorithm. We choose $\frac{1}{2}\qty(a+b)=1.75$ as a representative regime where neither the neutral or strong competition approximations apply. The numerical results are shown in Figure \ref{fig16}, together with the theoretical predictions in the binodal and spinodal limits derived in Secs.~\ref{sec:noneqbin} and \ref{sec:noneqspin}.

Close to the binodal, we use the prediction of Eq.~(\ref{eq:ucfit}) for $U_\mrm c$ with $\varepsilon=a-b$. The constant $K$ for Langevin noise, denoted $K_\mrm l$, is calculated using Eqs.~(\ref{eq:speed}), (\ref{eq:uc}) and (\ref{eq:deltaf}) using the profile of the front found numerically but independently of the data from the GMAM algorithm. The constant $\mathcal U$ is fitted using the points from the GMAM calculations close to the binodal. For $\frac{1}{2}\qty(a+b)=1.75$, we found $\mathcal U=-0.60$.

Figure \ref{fig16} shows excellent agreement between GMAM calculations and non-equilibrium nucleation theory predictions up to values of $a-b$ around $0.35$. This is consistent with theoretical estimates of their range of validity, which supposes $R_\mrm c\gtrsim w$. At $\frac{1}{2}\qty(a+b)=1.75$, we find from Figs.~(\ref{fig11}, \ref{fig13}) a speed $c\approx0.4\qty(a-b)\sqrt{D}$ and $w\approx7\sqrt{D}$, so that using Eq.~(\ref{eq:rc}), our predictions should indeed be valid when $a-b\lesssim0.35$.

\begin{figure}
    \centering
    \includegraphics[width=0.9\linewidth]{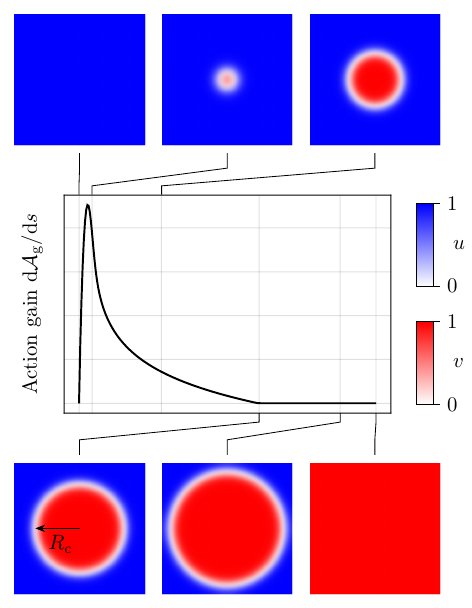}
    \caption{MAP for the LV system with parameters $a=1.8$, $b=1.7$, $D=1$ as well as its gain in action $\mrm d\mathcal A_\mrm g/\mrm ds$ as a function of the position along the path $s$. The MAP was obtained using the numerical procedure described in section \ref{sec:num}.}
    \label{fig17}
\end{figure}

Furthermore, we show the MAP found numerically for $a-b=0.2$ in Fig.~\ref{fig17}, as well as the geometric action $A_\mrm g$ gained along the path. The MAP consists in a stage of front creation, followed by their propagation over the whole space. No action is gained beyond the critical nucleus because the MAP follows the deterministic path. This picture confirms the assumptions made to derive the theory close to the binodal. 

In the spinodal limit, the quasipotential in two dimensions is, according to Eq.~(\ref{eq:ucspin}),
\begin{equation}
    U_\mrm c=D\qty(\frac{b-1}{a-1})^2\mathcal E_\mrm{sp}\qty[\phi_\mrm c]\text.\label{eq:ucspinlv}
\end{equation}
The range of validity of this equation, as hatched in Fig.~\ref{fig14}(c), implies in particular that there is no crossover regime between spinodal and binodal nucleation, which involve qualitatively different critical nuclei and MAPs. Nevertheless, Fig.~\ref{fig16} shows excellent agreement between the spinodal theory and GMAM calculations over a large range of $a-b$. Combining the two analysis one can obtain an accurate description over the entire domain of $a-b$. This indicates \textit{a posteriori} that the spinodal condition $b-1\ll\min\qty(1,\,1-1/a)$ is not as restrictive as it may appear.

Hence, GMAM computations and theoretical predictions agree on $U_\mrm c$ at fixed $\frac{1}{2}\qty(a+b)$ and varying $a-b$, validating our theory for Langevin noise. We now use these results to investigate which ecological conditions make nucleation most likely.

\subsubsection{Ecological Discussion}
We now vary the competition strength $\frac{1}{2}\qty(a+b)$ along the binodal. The coefficient $K_\mrm l$ governing the nucleation rate, see Eq. (\ref{eq:ucfit}), is a joint function of the susceptibility and of the mobility for Langevin noise $g_\mrm l$, which in dimension two reads
\begin{equation}
    K_\mrm l=\pi\frac{D^2}{g_\mrm l\chi}\text.\label{eq:K}
\end{equation}

\begin{figure}
    \centering
    \includegraphics[width=\linewidth]{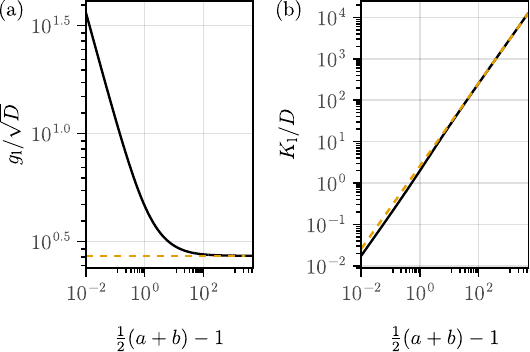}
    \caption{(a) Mobility $g_\mrm l$ and (b) coefficient $K_\mrm l$ for Langevin noise as a function of competition strength. Both ratios $g_\mrm l/\sqrt{D}$ and $K_\mrm l/D$ are independent of $D$ in dimension two. The black line in (a) is evaluated numerically from Eq.~(\ref{eq:mobility}), while the dashed line is the strong-competition asymptote, Eq.~(\ref{eq:mobilitylangevinstrong}). Both black and dashed lines in (b) are the result of combining the corresponding lines in (a) and Fig.~\ref{fig13} through Eq.~(\ref{eq:K}).}
    \label{fig18}
\end{figure}

The mobility for Langevin noise, $g_\mrm l$, is plotted as a function of $\frac{1}{2}\qty(a+b)$ in Fig.~\ref{fig18}(a). Because Langevin noise is independent of $\mb u$, its expression Eq.~(\ref{eq:mobility}) indicates that the front is susceptible to noise over its whole width, with a response controlled by $\abs{\boldsymbol{\delta}\mb u_0}^2$. The mobility therefore tracks the width of the front, diverging near neutrality at $a,\,b\to1$ and saturating in the limit of strong competition, as confirmed analytically in Appendix~\ref{sec:fronts}.

Combining these results with those for the susceptibility yields the coefficient $K_\mrm l$, shown in Fig.~\ref{fig18}(b). Together with the analytical results of Appendix~\ref{sec:fronts} in the limit of strong competition, Eq.~(\ref{eq:K}) shows that $K_\mrm l$ grows asymptotically linearly with $\frac{1}{2}\qty(a+b)-1$, and Fig.~\ref{fig18}(b) shows that this trend remains approximately valid over the whole range. Interestingly, the same linear growth arises from two different mechanisms. Near neutrality, the front has a single length scale, its width $w$, and both $\chi$ and $g_\mrm l$ are proportional to it, so that $K_\mrm l\propto1/w^2\propto\frac{1}{2}\qty(a+b)-1$. At strong competition, the width saturates and $g_\mrm l$ with it, so that the growth of $K_\mrm l$ comes entirely from the susceptibility, $\chi\propto1/\qty(a+b)$: it is the depletion region that makes nucleation harder.

These results mean that, at a given competitive advantage $a-b$, strongly competing systems have exponentially smaller nucleation probabilities than weakly competing ones. These findings follow directly from the structure of fronts, through both the susceptibility and the mobility, confirming that fronts are the link between local ecological interactions and nucleation. Langevin noise, however, is not realistic; we now show that the same trends survive when demographic noise is used instead.

\begin{figure}[t!]
    \centering
    \includegraphics[width=\linewidth]{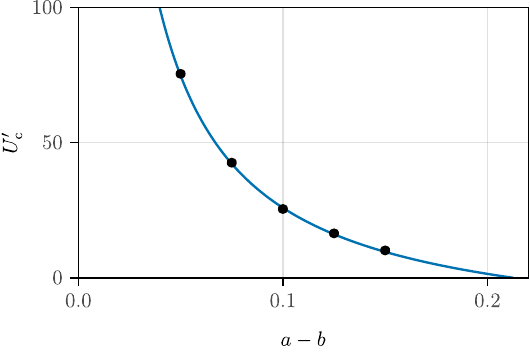}
    \caption{Quasipotential $U_\mrm c'$ for $\frac{1}{2}\qty(a+b)=1.75$, $D=1$ and varying $a-b$ away from the binodal, to the point where the critical nucleus has the same radius $R_0$ as the initial inoculum. Black dots are results from the GMAM and the blue line is the result of Eq.~(\ref{eq:ucdemo}) with $K_\mrm d$ evaluated numerically independently from the GMAM data and $\mathcal U'$ fitted to the GMAM data..}
    \label{fig19}
\end{figure}

\subsection{Quasipotential with Demographic Noise}
\label{sec:quasi}
Demographic noise is commonly used in ecology to model fluctuations in growth rates arising from finite and discrete population sizes~\cite{engen_demographic_1998}. Indeed, individual reproduction and death events occur at random times. Their effects average our in the large-population limit, but generate fluctuations a finite abundances. Demographic noise therefore has correlations proportional to the population size and is characterized by
\begin{equation}
    \mb G\qty(\mb u)=\mqty(u&0\\0&v)\text.
\end{equation}

As a model, demographic noise may break down at very small abundances, but it avoids the pathologies of Langevin noise. Specifically, it cannot lead to negative abundances, or generate a population from a state with zero individuals. 

The latter property has important consequences for nucleation: starting from a homogeneous state $\mb u_+^*$, nucleation is impossible because this state is absorbing under demographic noise. We therefore calculate the quasipotential for reaching $\mb u_-^*$ starting from a circular droplet of $\mb u_-^*$ embedded in an otherwise homogeneous state $\mb u_+^*$, and denote it by $U_\mrm c'$. We choose the initial radius $R_0=10\sqrt{D}$ independently of $a$ and $b$, slightly larger than the front width over the range of parameters considered. See Appendix \ref{sec:numpot} for implementation details.

As for Langevin noise, we first test our results at fixed $\frac{1}{2}\qty(a+b)=1.75$ while varying $a-b$, as shown in Fig.~\ref{fig19}. We use Eq.~(\ref{eq:ucfit}) as
\begin{equation}
    U_\mrm c'=\frac{K_\mrm d}{a-b}+\mathcal U'\qty(R_0)\label{eq:ucdemo}
\end{equation}
with
\begin{equation}
   K_\mrm d=\pi\frac{D^2}{g_\mrm d\chi}\text.\label{eq:Kdemo}
\end{equation}
For $R_0=10\sqrt{D}$, we fit $\mathcal U'\qty(R_0)=-23$. Figure \ref{fig19} shows excellent agreement between theory and GMAM calculations. Let us now vary $\frac{1}{2}\qty(a+b)$.

\begin{figure}
    \centering
    \includegraphics[width=\linewidth]{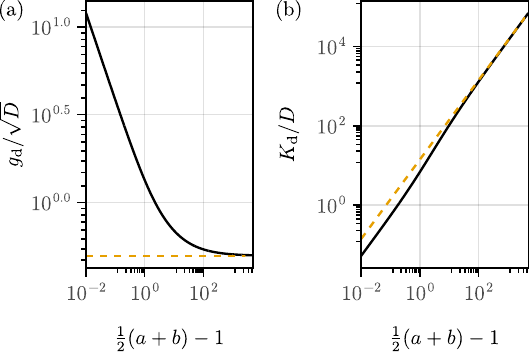}
    \caption{Equivalent of Fig.~\ref{fig18} for demographic noise. Both ratios $g_\mrm d/\sqrt{D}$ and $K_\mrm d/D$ are independent of $D$ in dimension two. The black line in (a) is evaluated numerically from Eq.~(\ref{eq:mobility}), while the dashed line is the strong-competition asymptote, Eq.~(\ref{eq:mobilitydemostrong}). Both black and dashed lines in (b) are the result of combining the corresponding lines in (a) and Fig.~\ref{fig13} through Eq.~(\ref{eq:Kdemo}).}
    \label{fig20}
\end{figure}

The mobility for demographic noise, $g_\mrm d$, is plotted as a function of $\frac{1}{2}\qty(a+b)$ in Fig.~\ref{fig20}(a). It is smaller than the mobility for Langevin noise $g_\mrm l$ because only the high-abundance regions of the front contribute significantly to $g_\mrm d$. However, this comparison is not directly meaningful since the noise amplitude $T$ may have different origins for the two noise models. Still, $g_\mrm d$ follows the same trends as $g_\mrm l$: it diverges near neutrality and approaches a constant at strong competition, reflecting the same underlying changes in front structure.

As for Langevin noise, combining $g_\mrm d$ with the susceptibility $\chi$ gives the coefficient $K_\mrm d$, shown in Fig.~\ref{fig20}(b). It follows the same trends as $K_\mrm l$, growing approximately linearly with $\frac{1}{2}\qty(a+b)-1$. The coefficients of proportionality in the strong-competition and neutral limits, however, differ more noticeably than with Langevin noise. Because demographic noise provides a more realistic description of ecosystems, these results allow us to appreciate the ecological significance of Fig.~\ref{fig20}(b): they determine the invasion probability of a small inoculum. Close to the binodal, the first term of Eq.~(\ref{eq:ucdemo}) dominates, so that in two dimensions
\begin{equation}
    \fbox{$\displaystyle\ln p\qty(R_0)\simeq-\frac{K_\mrm d}{T\qty(a-b)}\simeq-\Lambda\frac{D}{T}\frac{a+b-2}{a-b}$}\text,\label{eq:lnp}
\end{equation}
where $\Lambda$ is the dimensionless constant defined by the approximately linear behavior $K_\mrm d\simeq\Lambda D\qty(a+b-2)$ of Fig.~\ref{fig20}(b), for which we find $\Lambda=\frac{14\pi\qty(9-2\sqrt{3})^3}{225\qty(3+\sqrt{3})}\approx7.0$ (appendix \ref{sec:fronts}). This holds for any initial radius much smaller than the critical radius but larger than the front width $w$, as is the case for the choice $R_0=10\sqrt{D}$ used in our simulations. Thus, at fixed competitive advantage $a-b$, increasing the overall competition strength $a+b$ makes invasion {\it exponentially} less likely. The magnitude of the effect is set by $D/T$, which has a direct ecological meaning: it is  proportional to the local population size and the strength of dispersal. 

This prediction can be tested without adjustable parameters. In engineered microbial systems, the antagonism coefficients $a$ and $b$ are set by the production of, and the immunity to, toxins~\cite{giometto2021}: raising the toxin production of both strains together increases $a+b$ while leaving the asymmetry $a-b$ essentially unchanged, which is precisely the comparison Eq.~(\ref{eq:lnp}) describes. Both ingredients are measurable on the front itself: the width $w$ from the abundance profiles, and the critical radius from Eq.~(\ref{eq:rc}), $R_\mrm c=D/c$ in two dimensions. We show in the appendix, see Eqs.~(\ref{eq:frontneutralxi}, \ref{eq:speedstrong}), that $R_\mrm c/w\propto\qty(a+b-2)/\qty(a-b)$, so that the critical inoculum should grow linearly with respect to the width of fronts as the competition strength is increased at fixed competitive advantage, while the invasion probability of subcritical inocula falls exponentially.

Finally, the similarity between the results for Langevin and demographic noises suggests that these results may extend to other noise models, such as environmental noise, which represents coherent fluctuations in individual growth rates,
\begin{equation}
    \mb G\qty(\mb u)=\mqty(u^2&0\\0&v^2)\text,
\end{equation}
or to a combination of different noise sources, provided that they act across the entire front.

\section{Discussion And Perspectives\label{sec:conclusion}}
We have developed a theory of nucleation for non-equilibrium dynamics with non-conserved, vector order parameters, in two limits that require distinct perturbative treatments. Close to the binodal, the critical nucleus is a droplet bounded by well-formed fronts, whose inner profile we coarse-grain away; close to the spinodal, it is a wide perturbation of small amplitude. In the first limit, our central result is that the coarse-grained motion of a front obeys the same equation whether or not the underlying field dynamics is at equilibrium, so that nucleation is controlled by two quantities directly related to the front profile: its speed $c$, which takes over the conceptual role of the bulk free-energy difference of CNT, and its mobility $g$. We have checked these predictions against direct numerical calculations of the quasipotential in a bistable Lotka-Volterra model, and used them to characterize how nucleation rates follow from the strength of ecological interactions.

Cates and coworkers proposed in a concurrent work a theory of nucleation near the binodal for non-equilibrium dynamics with non-conserved, scalar order parameters~\cite{chatzittofi_nonequilibrium_2026, ziethen_nucleation_2026}. Our work shares with these studies, and with recent work on fluctuating interfaces~\cite{sarfati_bulk_2026}, the use of an element of the kernel of $\mathcal L_\mrm b^\dagger$ to define the position of the interface. There, this choice is introduced as a convenient one among several ways of locating a fluctuating interface; here it emerges as the solvability condition of a singular perturbation problem, which fixes it uniquely. 
None of these works treats vector order parameters, and it is the vector structure--the depletion region inside the front--that carries the ecological content of our results.

Beyond the two-species system studied here, our framework is relevant to disparate biological systems in which one ecological or chemical state invades another. In microbial communities, spatial invasion is at play in questions such as microbial antagonism~\cite{garcia-bayona_bacterial_2018}, antibiotic mediated switching~\cite{bucci2012social} and the establishment of transplanted microbiota~\cite{li_spatial_2020}. Our spatially extended description provides a natural setting in which to ask when an initially localized population can successfully invade its surroundings. Non-equilibrium reaction-diffusion models of biological condensates offer another class of systems in which nucleation must be described in the presence of coupled chemical fields~\cite{brangwynne_germline_2009, weber_physics_2019, brangwynne_phase_2013, zwicker_intertwined_2022, bauermann_chemical_2022}.

The vector nature of our order parameter is what makes such applications reachable. In microbial antagonism, for instance, the diffusion of toxins or resources strongly affects the abundance profiles across fronts~\cite{giometto2021}. Treating these fields explicitly captures their coupling to population dynamics, and the effects that an effective scalar description would discard. It would clarify how the structure and the sensitivity of fronts depend on competitive advantage and on noise in realistic microbial ecosystems. More generally, extending our framework to multi-species models such as the generalized Lotka--Volterra model could shed light on invasion in complex communities, and on its consequences for coexistence and ecosystem robustness~\cite{amor_transient_2020, al-hiyasat_spatiotemporal_2026}.

Our theory may also provide a general framework for studying nucleation-driven phenomena in non-reciprocal systems~\cite{fruchart_non-reciprocal_2021}. The non-reciprocal Ising model, which provides a natural non-equilibrium counterpart to the equilibrium Ising model underlying the classical picture of nucleation, illustrated in Fig.~\ref{fig1}, was recently shown to exhibit complex nucleation dynamics, including droplet-induced swaps and droplet capture through successive nucleation events~\cite{avni_nonreciprocal_2025, avni_dynamical_2025, schuttler_nonreciprocal_2025}. These phenomena were described using approximate, piecewise reductions to equilibrium dynamics. Our framework could provide a systematic approach to such genuinely non-equilibrium nucleation processes. More generally, as discussed in Sec.~\ref{sec:generalizations}, our theory extends straightforwardly to a broader class of non-equilibrium models beyond Eq.~(\ref{eq:rd}), including some exhibiting active phase separation.

Increasing the complexity of the local dynamics also raises a more fundamental challenge. In systems with many interacting species, with additional chemical fields, or with strong non-reciprocity, where oscillatory and traveling states are generic, the internal dynamics of a front may itself become nontrivial, with oscillations or even chaos. Fronts would then cease to be the rigid, gapped objects that our coarse-graining relies on. Building a coarse-grained description of such dynamically structured fronts, and understanding how their internal dynamics feeds back on nucleation and invasion, is certainly worth future studies.

\begin{acknowledgments}
We thank Eric Vanden-Eijnden for suggesting to initialize the GMAM from the result of the string method. We thank Daniel Amor, Andrea Giometto, and Elisa Garabello for discussions on their experiments of antagonistic bacterial strains. We also thank Michel Fruchart for clarifying the complex nucleation events in the non-reciprocal Ising model. GB acknowledges support from ANR grant LinkFM (ANR-25-CE30-4234).
\end{acknowledgments}

\section{Data Availability}

The data and code that support the findings of this article are openly available~\cite{zenodo}. 

\appendix

\section{Details of Numerical Methods\label{sec:detailnum}}

All calculations were performed with $D=1$ in the Julia programming language, version 1.11.7, on machines that ran Ubuntu 24.04.4.

\subsection{Fronts\label{sec:numfronts}}

Simulations of Eq.~(\ref{eq:rd}) were performed using an explicit Euler finite differences scheme with Neumann boundary conditions and a symmetric three-point stencil for the laplacian operator, using $N=2048$ spatial nodes and a time step of $\mrm dt = 10^{-4}$. The spatial length of the simulation window was chosen much greater than the front width, as low as $L=40$ for $a$ and $b$ large. For large values of $a$ and $b$, the solution converges to a uniformly translating profile on time scales of order one, which we held to be the front and extracted after times of order $100$. For $a,b$ close to one, convergence is slower. We monitored it using the global observable $\int v\qty(x)\mrm dx$.

We define the front position $x$ numerically by $u\qty(x)=v\qty(x)$. When simulating a front with nonzero speed, we reset the center of the front after each time step so that it remains at the center of the simulation window. This was performed by shifting the values of all nodes an amount equal to the distance the front had moved and filling the remaining values with $\mb u_\pm^*$, depending on the sign of $c$. Once convergence to the front is ensured, we estimate its speed $c$ by cumulating these distances over the simulation duration, of order $1000$.

For different values of $a$ and $b$, we investigated the effect of $N$ on the front profile and speed. We find that front profiles can be well resolved with as little as $10$ discretization points across their width. Front speed, however, is sensitive to $N$. Below some threshold, fronts are stationary in the simulation, when theory and more accurate simulations predict a nonzero speed. Slightly above that threshold, the front position oscillates as it propagates. For the production runs, we used large values of $N$, above the threshold, by checking that the front position increases linearly in time.

This threshold phenomenon could be relevant to experiments seeking to study front propagation in microbial communities by using grids of wells and explicitly implementing diffusion by mixing in neighboring wells via repeated dilution phase. Indeed, a well grid is similar to a finite differences mesh and can be expected to generate similar discretization effects. This phenomenon seems similar to a locking of fronts in periodic potentials relevant for hydrodynamic purposes~\cite{pomeau_front_1986}.

\subsection{Critical Nucleus\label{sec:numnuc}}

We use an explicit Euler scheme to implement the shrinking dimer algorithm. The field $\mb V$ in the SDA evolves with a timescale thrice as fast as $\mb u$, which empirically ensures good performance of the algorithm. Except for Figure \ref{fig15}, we use a square domain with size $L$ larger than the diameter of the critical radius plus several times the front width.

The original SDA of Ref.~\cite{zhang_shrinking_2012} does not use the Jacobian $\mb J\qty(\mb u)$, but an approximation of it based on finite differences of $\mb b$. We implemented the SDA method using $\mb J$ directly, because we can compute it and the LV flow $\mb b$ is smooth enough for purely local dynamics to converge well.

The initial conditions for $\mb u$ are the result of a direct simulation of Eq.~(\ref{eq:rd}) for a time span of $5$ units, whose own initial conditions are a disk of the stable state in the metastable one, such that $\mb u\qty(t=0,\,\abs{\mb x}\leq R)=\mb u_-^*$ and $\mb u\qty(t=0,\,\abs{\mb x}>R)=\mb u_+^*$. We find the radius $R$ by dichotomy to give a rough estimate of the critical radius $R_\mrm c$, based on the sign of the quantity $\dv{t}\int v\qty(\mb x)\mrm d\mb x$ after the edges of the disk relax to front-like profiles.

The initial conditions for $\mb V$ are proportional to the numerical estimate of the radial derivative of $\mb u$. Indeed, we expect the unstable mode of the critical nucleus to correspond to a change in $R$ close to the binodal. The coefficient of proportionality ensures normalization of $\mb V$.

These initial conditions provide a good enough estimate of the critical nucleus and its unstable mode, which ensures fast convergence of the algorithm. However, we also added noise to the initial conditions and tested convergence using values of $R$ far from $R_\mrm c$ and still found good performance of the shrinking dimer algorithm. In particular, this confirmed that the critical nucleus has spherical symmetry and allowed to simulate the finite-size effects of Figure \ref{fig15} using the same initial conditions as the large-$L$ case.

We use periodic boundary conditions, so that both the SDA and the direct simulations used to provide its initial conditions are implemented using a semi-implicit pseudo-spectral method. The force term $\mb f_\mb p$ and its Jacobian $\mb J_\mb p$ are calculated explicitly at each site in direct space, while the diffusion term is implemented implicitly in the Fourier basis. Real space is discretized on a $N^2=512\times512$ square lattice. Each time step of the SDA follows this pseudo-code:

\begin{itemize}
    \item calculate $A=\left\langle\mb V\qty(t),\,\mb u\qty(t)\right\rangle$\;;
    \item evolve $\mb u\qty(t)$ using the semi-implicit pseudo-spectral method for the flow $\mb b\qty(\mb  u)=D\Delta\mb u+\mb f_\mb p\qty(\mb u)$ and save the result as $\mb u\qty(t+\mrm dt)$\;;
    \item add the quantity $2\qty(A-\left\langle\mb V\qty(t),\,\mb u\qty(t+\mrm dt)\right\rangle)\mb V\qty(t)$ to $\mb u\qty(t+\mrm dt)$\;;
    \item step $\mb V\qty(t)$ using the semi-implicit pseudo-spectral method for the flow $\mb J\qty(\mb u)\mb V=D\Delta\mb V+\mb J_\mb p\qty(\mb u)\cdot\mb V$ and save the result as $\mb V\qty(t+\mrm dt)$\;;
    \item normalize $\mb V\qty(t+\mrm dt)$\text.
\end{itemize}

We use a large timestep $\mrm dt = 0.3$ since the algorithm is only used to provide the field $\mb u$ after a long time, and the transient dynamics is of no relevance. The number of time steps, $2\times10^4$, ensures convergence of the observable $\int v\qty(\mb x)\mrm dx$. We also checked visually the convergence of the critical nucleus profile.

In order to calculate the universal spinodal nucleus $\psi_\mrm c$, we apply the SDA to the equation
\begin{equation}
    \frac{\partial\psi}{\partial t}=\Delta\psi-\psi+\psi^2\text.
\end{equation}
Indeed, $\psi_\mrm c$ is a critical nucleus between the states $\psi=0$ and $\psi=\infty$. As such, it is a saddle point with one unstable dimension, so that the SDA can be used. We choose initial conditions that approximately mimic the expected shape for $\psi_\mrm c$, namely
\begin{equation}
    \psi\qty(r)=1.8\exp\qty(-\max\qty(0,\,r-1.5))\text.
\end{equation}

\subsection{Quasipotential\label{sec:numpot}}

Each image was simulated with periodic boundary conditions and a semi-implicit pseudo-spectral method on a square lattice of size $N^2$. For most simulations, $N=512$, especially close to the binodal where the critical nucleus is large, as explained in Sec.~\ref{sec:numgmam}. For simulations further away the binodal, we use $N$ as low as $64$. For most simulations, we used $N_s=400$ images for the path. Away from the binodal, $N_s$ was taken as low as $100$. Some simulations were performed independently using $\qty(N=512,\,N_s=400)$ and $\qty(N=64,\,N_s=100)$ and values we found for the quasipotential had a difference smaller than the size of points in Figure \ref{fig16}.

As in the SDA, we choose large time steps to speed up convergence since the transient dynamics is of no interest. The string method consists of $10$ sets of $30$ time steps of evolution of each image during $\mrm dt=0.5$. Each set is followed by a linear interpolation step. We perform a number of steps ranging from $4000$ to $9000$, with a step size as large as $0.5$ to ensure numerical stability. Convergence is monitored by checking that the action is stationary.

The evolution step of the GMAM algorithm is implemented as
\begin{widetext}
    \begin{multline}
        \Tilde{u}_i\qty(t,\,s,\,\mb x)=\mrm dt\lambda^2\frac{\partial^2\Tilde{u}_i}{\partial s^2}-\mrm dtD^2\Delta^2\Tilde{u}_i+u_i+\mrm dt\qty{-\lambda\partial_jf_{\mb pi}\frac{\partial u_j}{\partial s}+\lambda\frac{\partial\lambda}{\partial s}\frac{\partial u_i}{\partial s}-D\Delta f_{\mb pi}}\\
        +\mrm dt\qty{G_{ij}\partial_jf_{\mb pk}\qty(\mb G^{-1})_{kl}\qty(\lambda\frac{\partial u_l}{\partial s}-b_l)-\frac{1}{2}G_{ij}\partial_j\qty(\mb G^{-1})_{kl}\qty(b_k-\lambda\frac{\partial u_k}{\partial s})\qty(b_l-\lambda\frac{\partial u_l}{\partial s})}\label{eq:gmam}
    \end{multline}
\end{widetext}
where $\lambda\qty(t,\,s)=\norm{\mb G^{-\frac{1}{2}}\cdot\mb b}/\norm{\mb G^{-\frac{1}{2}}\cdot\partial\mb u/\partial s}$ and $\mb b\qty(t,\,s,\,\mb x)=D\Delta\mb u+\mb f_\mb p$.

In Eq.~(\ref{eq:gmam}), all terms involving $\Tilde{u}_i$ are treated implicitly using the Thomas algorithm and both $\Delta^2$ and all terms involving $\Delta$ are calculated in Fourier space. All first order derivatives in $s$ are calculated using a symmetric two-point finite-difference stencil.

For demographic noise, $\mb G$ needs to be regularized so that its inverse does not diverge when one species reaches low abundances, so we use
\begin{equation}
    \mb G\qty(\mb u)=\mqty(\abs{u+\mrm{reg}}&0\\0&\abs{v+\mrm{reg}})\text,
\end{equation}
and employ $\mrm{reg}=10^{-4}$. We tested values of $\mrm{reg}$ as low as $10^{-10}$ and observed no difference on the results for the quasipotential. To calculate $U_\mrm c'$, we first find the MAP between $\mb u_+^*$ and $\mb u_\mrm c$, then the point $\mb u_0$ along this path which has the closest radius to $R_0$. For this, we choose the radius as the point where $u=v$. The action is then integrated from $\mb u_0$ to $\mb u_\mrm c$. 

\section{Analytical Study of Fronts\label{sec:fronts}}

We present an analytical study of fronts in the LV system in two tractable limits along the binodal line: neutral and strong competition. We perform the change of variables $A=u_\mrm{fr}+v_\mrm{fr}$, $\xi=\qty(u_\mrm{fr}-v_\mrm{fr})/A$ so that Eq.~(\ref{eq:front}) becomes
\begin{equation}
    -c\dv{A}{x}=D\dv[2]{A}{x}+A\qty(1-A)-\qty(a+b-2)A^2\frac{1-\xi^2}{4}\label{eq:frontA}
\end{equation}
and
\begin{eqnarray}
    -cA\dv{\xi}{x}&=&D\qty(2\dv{A}{x}\dv{\xi}{x}+A\dv[2]{\xi}{x})\nonumber\\
    &&+\qty[\qty(a+b-2)\xi-\qty(a-b)]A^2\frac{1-\xi^2}{4}\text.\label{eq:frontdelta}
\end{eqnarray}

\subsection{Neutral Limit}

As $a,b\to1$, Eq.~(\ref{eq:frontA}) simplifies to the equation for Fisher fronts
\begin{equation}
    -c\dv{A}{x}=D\dv[2]{A}{x}+A\qty(1-A)
\end{equation}
with boundary conditions $A\qty(x\to\pm\infty)=1$ and under the constraint $A\qty(x)\geqslant0$, which only has the solution $A\qty(x)=1$. Then, Eq.~(\ref{eq:frontdelta}) can be solved by~\cite{mckean_nagumos_1970}
\begin{equation}
    \xi\qty(x)=\tanh\qty(\frac{x}{w}),\quad w=\sqrt{\frac{8D}{a+b-2}},\quad c=w\frac{a-b}{4}\text.\label{eq:frontneutralxi}
\end{equation}

This gives the leading order behavior as $a,\,b\to1$. A correction to $A\qty(x)$ can be obtained self-consistently by linearizing Eq.~(\ref{eq:frontA}) around $A=1$ and injecting Eq.~(\ref{eq:frontneutralxi}), yielding
\begin{equation}
    A\qty(x)=1-\frac{a+b-2}{4\cosh^2\qty(x/w)}\text.\label{eq:frontneutralA}
\end{equation}

This asymptotic form matches very well with the fronts found by numerically simulating Eq.~(\ref{eq:rd}), see right panel of Fig.~\ref{fig12}.

\subsection{Strong Competition Limit}

\subsubsection{Shape of Fronts}

When $a=b\to+\infty$, it may be seen self-consistently that $\xi$ varies on much smaller scales than $A$. In fact, $\xi$ varies on scales $O\qty(\sqrt{D/\qty(a+b-2)})$ as above. Therefore, in Eq.~(\ref{eq:frontA}), $\xi^2\qty(x)\approx\xi^2\qty(\pm\infty)\approx1$ almost everywhere, except in a thin region around $x=0$, where $\xi\qty(0)=0$. However, $a+b-2$ is very large, such that the term $\qty(a+b-2)A^2\qty(1-\xi^2)/4$ can be replaced by a delta function with a certain amplitude $B$. Since $a=b$, we have $c=0$ leading to
\begin{equation}
    0=D\dv[2]{A}{x}+A\qty(1-A)-B\delta\qty(x)\text.\label{eq:eqfrontstrongA}
\end{equation}

Solutions of this equation for $x\gtrless0$ verifying $A\qty(\pm\infty)=1$ are of the form
\begin{equation}
    A\qty(x)=\frac{3}{2}\tanh^2\qty(\pm\frac{x-x_0}{w})-\frac{1}{2},\quad w=2\sqrt{D}\text,\label{eq:frontstrongw}
\end{equation}
which can be patched to give an even function as
\begin{equation}
    A\qty(x)=\frac{3}{2}\tanh^2\qty(\frac{\abs{x}}{w}+C)-\frac{1}{2}\text,\label{eq:frontstrongA}
\end{equation}
where $C$ is a constant. In the limit $a=b\to+\infty$, $A\qty(0)$ should tend towards $0$, yielding $C=\operatorname{arctanh}\qty(1/\sqrt{3})$.

Comparison between this function and numerical results for large $a=b=50$ is shown in the right panel of Fig.~\ref{fig12}. The region of mismatch is the support of $1-\xi^2$ and its size shrinks as $O\qty(\sqrt{D/\qty(a+b-2)})=O\qty(\sqrt{D/a})$.

\subsubsection{Linear Response Of Fronts}

The Jacobian of $\mb f$, expressed in terms of $A$ and $\xi$, is
\begin{align}
    \mb J&=\mqty(1-2u-av&-au\\-bv&1-2v-bu)\\
    &=\mathbbm{1}-A\mqty(1+\xi+a\frac{1-\xi}{2}&a\frac{1+\xi}{2}\\b\frac{1-\xi}{2}&1-\xi+b\frac{1+\xi}{2})\text.
\end{align}

In the limit $a=b\to+\infty$ and on scales much larger than $\sqrt{D/a}$, we write its transpose as
\begin{equation}
    \mb J^\mathsf T=-Aa\mqty(\Theta\qty(-x)&\Theta\qty(-x)\\\Theta\qty(x)&\Theta\qty(x))+O\qty(1)\text.
\end{equation}

This matrix has eigenvalues $O\qty(1)$ and $-Aa\qty(\Theta\qty(x)+\Theta\qty(-x))+O\qty(1)=-Aa+O\qty(1)$, associated with the eigenvectors $\mqty(-1\\1)+o\qty(1)$ and $\mqty(\Theta\qty(-x)\\-\Theta\qty(x))+o\qty(1)$ respectively. By definition, $\boldsymbol{\delta}\mb u_0$ verifies
\begin{equation}
    \mathcal L^\dagger\qty[\boldsymbol{\delta}\mb u_0]=D\dv[2]{\boldsymbol{\delta}\mb u_0}{x}+\mb J^\mathsf T\qty(\mb u_\mrm{fr}^\mrm b)\cdot\boldsymbol{\delta}\mb u_0=\mb0\text.
\end{equation}

For $\abs{x}\gg\sqrt{D/a}$, $\boldsymbol{\delta}\mb u_0$ thus concentrates on the eigenvector of $\mb J^\mathsf T$ associated with its lower eigenvalue, namely $\mqty(-1\\1)$. In the limit $a\to+\infty$, $\boldsymbol{\delta}\mb u_0$ converges to $\delta u_0\qty(x)\mqty(-1\\1)$ for all $x$, where we find that $\delta u_0\qty(x)$ is any solution to
\begin{equation}
    D\dv[2]{\delta u_0}{x}+\qty(1-2A\qty(x))\delta u_0=0
\end{equation}
that goes to $0$ at $x\to\pm\infty$. It can be checked explicitly, in analogy with the calculation leading to $\mathcal L\qty[-\mrm d\mb u_\mrm{fr0}/\mrm dx]=\mb0$, that one such solution is
\begin{equation}
    \delta u_0\qty(x)=\abs{\dv{A}{x}}=\frac{3}{w}\frac{\sinh\qty(\frac{\abs{x}}{w}+C)}{\cosh^3\qty(\frac{\abs{x}}{w}+C)}
\end{equation}
with $w=2\sqrt{D}$ and $C=\operatorname{arctanh}\qty(1/\sqrt{3})$.

It is now a matter of straightforward integration to calculate $\Tilde{\gamma}$ as
\begin{align}
    \Tilde{\gamma}&=\left\langle\boldsymbol{\delta}\mb u_0,\,-\partial\mb u_\mrm{fr}^\mrm b/\partial x\right\rangle\\
    &=\int_{-\infty}^{+\infty}\abs{\dv{A}{x}}\mqty(1\\-1)\cdot\dv{x}\mqty(A\Theta\qty(x)\\A\Theta\qty(-x))\mrm dx\\
    &=\int_{-\infty}^{+\infty}\qty(\dv{A}{x})^2+\abs{\dv{A}{x}}A2\delta\qty(x)\mrm dx\\
    &=2\int_0^{+\infty}\frac{9}{w^2}\frac{\operatorname{sh}^2\qty(\frac{x}{w}+C)}{\operatorname{ch}^6\qty(\frac{x}{w}+C)}\mrm dx+2\dv{A}{x}\qty(0^+)A\qty(0)\\
    &=\frac{18}{w}\frac{2}{135}\qty(9-2\sqrt{3})=\frac{2}{15\sqrt{D}}\qty(9-2\sqrt{3})\text.
\end{align}

Take now $a, b\to+\infty$ but $a\neq b$. Then, $c$ can be asymptotically calculated according to Eq.~(\ref{eq:speed}) as
\begin{align}
    \Tilde{\gamma}c&=\left\langle\boldsymbol{\delta}\mb u_0,\,\Delta\mb f\qty(\mb u_\mrm{fr}^\mrm b)\right\rangle\\
    &=\int_{-\infty}^{+\infty}\delta u_0\qty(x)\mqty(-1\\1)\cdot\qty[\qty(a-b)\frac{u_\mrm{fr}\qty(x)v_\mrm{fr}\qty(x)}{2}\mqty(-1\\1)]\mrm dx\\
    &=\qty(a-b)\delta u_0\qty(0)\int_{-\infty}^{+\infty}u_\mrm{fr}\qty(x)v_\mrm{fr}\qty(x)\mrm dx\qty(1+O\qty(\frac{1}{a}))\label{eq:casympstrong0}
\end{align}
since $u_\mrm{fr}\qty(x)v_\mrm{fr}\qty(x)$ is only meaningfully non-zero on scales $O\qty(\sqrt{D/\qty(a+b)})$ while $\delta u_0$ varies on scales $O\qty(\sqrt{D})$.

To get $c$ at leading order in $a-b$, $u_\mrm{fr}\qty(x)v_\mrm{fr}\qty(x)$ can be evaluated at $a=b$. Now, $uv=A^2\qty(1-\xi^2)/4$, so using Eq.~(\ref{eq:frontA}),
\begin{align}
    \int_{-\infty}^{+\infty}&u_\mrm{fr}\qty(x)v_\mrm{fr}\qty(x)\mrm dx\nonumber\\
    &=\frac{1}{a+b-2}\int_{-\infty}^{+\infty}D\dv[2]{A}{x}+A\qty(1-A)\mrm dx\text.
\end{align}

Under the approximation of Eq.~(\ref{eq:eqfrontstrongA}), the limits of the integral can be taken from $0^-$ to $0^+$, yielding
\begin{align}
    \int_{-\infty}^{+\infty}u_\mrm{fr}\qty(x)v_\mrm{fr}\qty(x)\mrm dx&=\frac{1}{a+b-2}D\qty[\dv{A}{x}]_{0^-}^{0^+}\\
    &=\frac{2D}{a+b-2}\frac{2}{w\sqrt{3}}\text.
\end{align}

Finally, Eq.~(\ref{eq:casympstrong0}) can be simplified to get
\begin{equation}
    c=\frac{\sqrt{D}\qty(a-b)}{a+b-2}\frac{5}{9-2\sqrt{3}}\label{eq:speedstrong}
\end{equation}
at leading order in $a-b$ and $1/\qty(a+b-2)$.

The mobility for Langevin noise $\mb G\qty(\mb u)=\mb 1$ can also be computed in a straightforward way using Eq.~(\ref{eq:mobility}) as
\begin{equation}
    g_\mrm l=\left\langle\boldsymbol{\delta}\mb u_0,\,\boldsymbol{\delta}\mb u_0\right\rangle/\Tilde{\gamma}^2=\frac{15\sqrt{D}}{9-2\sqrt{3}}\text,\label{eq:mobilitylangevinstrong}
\end{equation}
and that for demographic noise as
\begin{equation}
    g_\mrm d=\left\langle\boldsymbol{\delta}\mb u_0,\,\mqty(u_\mrm{fr}&0\\0&v_\mrm{fr})\boldsymbol{\delta}\mb u_0\right\rangle/\Tilde{\gamma}^2=\frac{45\qty(3+\sqrt{3})\sqrt{D}}{14\qty(9-2\sqrt{3})^2}\text.\label{eq:mobilitydemostrong}
\end{equation}

\bibliography{nucleation}

\end{document}